\documentclass[]{aa}

\usepackage[utf8]{inputenc}
\usepackage{graphicx}
\usepackage[varg]{txfonts}

\bibpunct{(}{)}{;}{a}{}{,}
\newcommand{\xmm}{XMM-\textit{Newton}}
\newcommand{\ch}{\textit{Chandra}}
\newcommand{\nus}{\textit{NuSTAR}}
\newcommand{\om}{XMM-OM}
\newcommand{\xr}{\textit{Swift}-XRT}
\newcommand{\uv}{\textit{Swift}-UVOT}

\begin{document}
\title{Interpreting the long-term variability of the changing-look AGN Mrk 1018}

\author{S. Veronese\inst{\ref{astron},\ref{kapteyn}}
\and C. Vignali\inst{\ref{unibo},\ref{inaf}}
\and P. Severgnini\inst{\ref{inafmi}}
\and G. A. Matzeu\inst{\ref{unibo},\ref{inaf},\ref{esa}}
\and M. Cignoni\inst{\ref{inaf},\ref{unipi},\ref{infn}}}

\institute{Netherlands Institute for Radio Astronomy (ASTRON), Postbus 2, 7990 AA Dwingeloo, The Netherlands, \email{veronese@astron.nl}\label{astron}
\and Kapteyn Astronomical Institute, University of Groningen, PO Box 800, 9700 AV Groningen, The Netherlands\label{kapteyn}
\and Dipartimento di Fisica e Astronomia Augusto Righi, Università degli Studi di Bologna, via Gobetti 93/2, 40129 Bologna, Italy\label{unibo}
\and INAF - Osservatorio di Astrofisica e Scienza dello Spazio di Bologna, via Gobetti 93/3, 40129 Bologna, Italy\label{inaf}
\and INAF - Osservatorio Astronomico di Brera, Via Brera 28, 20121 Milano, Italy\label{inafmi}
\and Quasar Science Resources SL for ESA, European Space Astronomy Centre (ESAC), Science Operations Department, 28692, Villanueva de la Ca\~{n}ada, Madrid, Spain\label{esa}
\and Department of Physics - University of Pisa, Largo B. Pontecorvo 3, 56127, Pisa, Italy\label{unipi}
\and INFN, Largo B. Pontecorvo 3, 56127, Pisa, Italy\label{infn}}

\date{Received 28 September 2023 / Accepted 11 December 2023}

\abstract{We present a comprehensive study of the Changing-Look Active Galactic Nucleus (CL-AGN) Mrk 1018, employing the largest dataset of optical, UV, and X-ray spectro-photometric data ever assembled for this source. Our investigation focuses on a detailed analysis of X-ray spectra, broad-band photometry, and optical-to-X-ray spectral energy distribution (SED) fitting, aiming to unravel the nature of the changing-look behaviour observed in Mrk 1018 between 2005 and 2019.\\Through our analysis, we confirmed that in those 14 years the X-ray from the source underwent a significant spectral variation, with the hardness ratio between the 0.5-2 keV and the 2-10 keV band increasing from 0.2 $\pm$ 0.1 to 0.4 $\pm$ 0.1. We also validated the dramatic broad-band dimming, with the optical, UV and X-ray luminosities decreasing by a factor $>7$, $>24$ and $\sim9$, respectively. We found that the declining UV emission is driving these drops.\\By describing the X-ray spectra with a two-Comptonization model, with a hot ($kT\sim$ 100 keV) and a warm ($kT<1$ keV) comptonizing medium reprocessing the photons from the accretion disk, we reached the conclusion that between 2005 and 2019 the properties of the hot medium remained the same, while the warm component cooled down from a temperature of $\sim0.4$ keV to $\sim0.2$ keV. This cooling can be explained via the weakening of the magnetic fields in the accretion disk and is also the source of the UV dimming.\\We proposed that this decline is caused by the formation of a jet, in turn originated from the change of the state of the inner accretion flow from a geometrically thin, optically thick to a geometrically thick, optically thin structure. Our optical-to-X-ray SED fitting seems to support this conclusion, as the estimated accretion rate normalized to the Eddington rate in the bright state ($\mu\sim0.06$) is above the critical value $\mu=0.02$ for a stable radiative flow, while in the faint state we found $\mu\sim0.01<0.02$, compatible with an advective accretion. Instabilities arising at the interface of the state-transition are then able to reduce the viscous timescale from $\sim10^5$ years to the observed $\sim10$ years of Mrk 1018 variability, reconciling all the observational properties of this CL-AGN into a complex but elegant physically motivated framework.\\We explored, in the end, a possible mechanism triggering the state-transition of the inner accretion flow. Our speculation is that gaseous clouds are pushed onto the innermost regions of the AGN by a galactic (dynamical friction) and/or an extragalactic process (wet merger, cold chaotic accretion). When one of these cloud passes by the accretion disk, it deposits material onto it, causing the accretion flow to puff up and establishing the state-transition.\\If this scenario will be confirmed by future numerical simulations, it will open a new branch of study to place CL-AGN into our current understanding of the feeding and feedback of AGN. We also think that our results can be applied to other CL-AGN as well, wondering if an accretion rate of $\mu\sim0.02$, coupled with minor `disturbances' in the accretion disk, is indeed the primary factor prompting the complex changing-look phenomenon.}

\maketitle

\section{Introduction}\label{sec:intro}
Active Galactic Nuclei (AGN) are generally classified on the presence of both broad ($> 10^3$ km/s) and narrow optical emission lines (type 1) or on the presence of only the narrow component (type 2). According to the AGN unification model \citep{antonucci93}, the difference between the two classes is only due to the line of sight: in type 2 AGN, the dusty torus surrounding the SMBH at scales of pc-tens of pc obscures the broad lines emission, while in type 1 AGN the observer has a direct line of sight of the Broad Lines Region (BLR).\\However, the distinction between type 1 and type 2 cannot be a mere orientation effect, as there are evidences for AGN whose optical emission lines vary significantly with the appearing/disappearing of their broad component on timescale that goes from days to decades \citep{risaliti07,mcelroy16,yang18,macleod19,graham20}. They are the so-called Changing-Look AGN (CL-AGN) and are currently an open issue in AGN physics. Indeed, the variation might be related to changes in the column density \citep{risaliti05,marchese12,dexter19b}, i.e. eclipsing events, disk instabilities \citep{gezari17,dexter19b} or accretion rate variations \citep{lamassa15,ruan16}. \\Mrk 1018 (see Table \ref{table:obj} for the main observational properties) is currently one of the most studied CL-AGN \citep{mcelroy16,husemann16,krumpe17,lamassa17,kim18,noda18,dexter19a,hutsem20,feng21,lyu21,walsh23,brogan23} because of its peculiar behaviour. It was optically classified as a type 2 AGN with narrow optical emission lines in 1981 \citep{osterbrock81}, but later studies carried out by \citet{cohen86} and \citet{goodrich89} revealed that broad emission lines appear in the optical spectrum, coupled with an increase of H$\beta$ flux by a factor of 6.6. Therefore, Mrk 1018 was reclassified as type 1 AGN.\\While \citet{cohen86} concluded that the variability was the result of a brightening of the UV continuum, which had ionized more material, the spectropolarimetry measurements used by \citet{goodrich89} suggested that the disappearing of dusty material along the line of sight was the reason behind the spectral variation. No further studies in this regard have been performed until 2016, when the Close AGN Reference Survey \citep{mcelroy16,husemann16} discovered that the broad optical emission lines have almost disappeared by comparing the spectrum of Mrk 1018 taken by Sloan Digital Sky Survey (SDSS) in 2003 with the one of the Multi-Unit Spectroscopic Explorer (MUSE) in 2015. The spectral changes were accompanied by a dimming of the source, with the $r$-band magnitude dropping from 15.2 $\pm$ 0.2 mag to 17.0 $\pm$ 0.2 mag over a timescale of twelve years.\\A number of hypotheses were proposed in the attempt of explain the mechanism behind the behaviour of Mrk 1018: a tidal disruption of a massive star \citep{mcelroy16}; the transit of obscuring clouds \citep{mcelroy16,husemann16}; a progressively shutdown of the AGN activity due to a simple lack of fuel \citep{mcelroy16}; the presence of a binary SMBH system \citep{mcelroy16,husemann16}; a state-transition of the accretion disk \citep{noda18,dexter19a,lyu21}; a recoiling SMBH \citep{kim18} and a magnetically driven outflow in the accretion disk \citep{feng21}. Yet, no conclusive explanation of its peculiar behaviour has been found.\\The main outcome of these works is that the mechanism leading to the variability of Mrk 1018 is very likely residing in the innermost regions of the AGN, i.e., the accretion disk. However, the lack of a sufficient spectral coverage, statistics or temporal coverage was always limiting their interpretation, and the phenomenon triggering whatever perturbation is affecting the accretion disk is still an open question. Motivated by these considerations, we decided to analyse all the available X-ray and optical/UV data, because the radiation in these bands sample the accretion flow, in the attempt to understand what is making Mrk 1018 a CL-AGN.\\In Sec. \ref{sec:obs} we list the main properties of the observations used in this paper and the data reduction process of the raw archival data to obtain scientific-quality products. In Sec.\ref{sec:ana} we show the analysis performed to address the question on what is triggering the variability of Mrk 1018. In Sec. \ref{sec:disc} the main results will be discussed. In Sec. \ref{sec:scenario} we present our interpretation for the changing-look nature of Mrk 1018. Finally, a summary of our research is provided in Sec. \ref{sec:conc} and some future prospects are given in Sec. \ref{sec:future}.\\Throughout the paper we assume a flat cosmology with H\textsubscript{0} $=$ 70 km/s/Mpc and $\Omega_M=$ 0.3 \citep{planck20,brout22}. All the uncertainties are at the 90\% confidence level for one parameter of interest \citep{avni76} if not explicitly stated otherwise.

\begin{table}
    \caption{Mrk 1018 main observational properties.}
    \label{table:obj}
    \centering
    \begin{tabular}{c c c}
    \hline\hline
    Parameter & Mrk 1018 & References \\
    \hline
    R.A. (J200) [hh mm ss]                         & 02 06 16.0                 & (1) \\
    DEC. (J200) [deg arcmin arcsec]                & -00 17 29.2                & (1) \\
    Redshift                                       & 0.043                      & (2) \\
    Luminosity distance [Mpc]                      & 190.2                      & \\
    Scale length [kpc/arcsec]                      & 0.848                      & \\
    Morphology                                     & S0                         & \\
    Stellar Mass [M\textsubscript{$\odot$}]        & $8.3\times10^{10}$    & (3)\\
    M\textsubscript{BH} [M\textsubscript{$\odot$}] & $7\times10^7$ & (4)\\
    \hline
    \end{tabular}
    \tablefoot{A flat universe with H\textsubscript{0} $=$ 70 km/s/Mpc and $\Omega_M=$ 0.3 is assumed. M\textsubscript{BH} is derived from a broad-band SED fitting \citep{jin12b}, while the stellar mass was calculated using models for stellar populations and photoionization \citep{koss11}.}
    \tablebib{(1) \citet{gaiadr3}; (2) \citet{sdssdr9}; (3) \citet{koss11}; (4) \citet{jin12b}.}
\end{table}

\section{Observations and data reduction}\label{sec:obs}
The observational campaign presented in this paper comprised eighty-two datasets from four X-ray observatories (\ch{}, \xmm{}, \xr{}\footnote{The observation 00049654004 was omitted because data were taken whilst the spacecraft or instruments were in an anomalous state (see https://www.swift.ac.uk/support/anomaly.php), while 00035776038 was excluded because the data were faulty.} and \nus{}), for a total of $\sim$ 600 kiloseconds (ks) of exposure time. The sample comprises eight observations taken during the bright state, that correspond to the period 2005-2012 hereafter, and the remaining were taken during the faint state from 2012 to 2020. To overcome the limitations in the statistics, we combined most of the observations as described in the following sections. A summary of the final dataset that has been used in this study can be found in Table \ref{table:obs}.

\subsection{Common data reduction steps}\label{sec:red}
The data were reduced according to the following standard procedure:
\begin{itemize}
    \item data reprocessing, where the most recent calibrations are applied; bad pixel, chip gaps and Good Time Interval (GTI) are obtained; exposure maps, Ancillary Response Functions (ARF) and Response Matrix Functions (RMF) are made; images of the object are reconstructed;
    \item definition of source and background extraction regions;
    \item spectral binning (or grouping), i.e., channels are grouped to have more than 20 counts per energy bins (necessary to apply a $\chi^2$ statistics).
\end{itemize}
Details for each instrument are provided in the following.

\subsection{\ch{} data reduction}\label{sec:chred}
\ch{} \citep{chandra} observed Mrk 1018 eleven times: one in 2010, one in 2016 and nine in the 2017-2019 period. The exposure time varies from 18 ks to 51 ks, with a mean value of 26 ks. Its bandwidth (0.3-7 keV) covers the energy range where we expected to find the bulk of the X-ray emission.\\The data were reduced using tools from CIAO \citep{ciao} v4.12 package. The calibration parameters during the reprocessing steps were automatically retrieved by CIAO from the online HEASARC Calibration Database CALDB v4.9.0.\\The AGN spectrum was extracted for each observation from a circular region with radius of 2.5'', so that the Encircled Energy Fraction (EEF) will be $\sim$ 90\% at 5 keV. The background spectrum was taken from four circular sources-free regions far from Mrk 1018 (angular distances and radii were arbitrary).\\The architecture of \ch{} detector would potentially suffer from pileup effect, a faulty measurement of two or more photons detected as a single particle whose energy is the sum of the photon’s energy. We evaluated how much each dataset was affected by this issue with the online tool PIMMS v4.10\footnote{https://cxc.harvard.edu/toolkit/pimms.jsp} and found that for the observation with ID 12868 a high fraction of the data (ca. 23\%) was affected by this effect. In order to take care of that, we added a pileup component in the expected spectrum of this dataset as suggested by the previous work of \citet{lamassa17} over the same observation.\\Given the very low count rate ($C$) of the other datasets with respect to 12868, we have assumed that these observations were taken during the faint state. Thus, after ensuring that the count rate and the hardness ratio $\frac{C_{2-10\text{ keV}}-C_{0.5-2\text{ keV}}}{C_{2-10\text{ keV}}+C_{0.5-2\text{ keV}}}$ of these other datasets are comparable, we co-added all of them to form a combined spectrum with the task \textit{combine\_spectra}. The spectra have been finally grouped to have at least 30 counts per bin.

\subsection{\xmm{} data reduction}
\xmm{} \citep{xmm} observed Mrk 1018 for four times: in 2005, 2008, 2018 and 2019. The exposure time varies from 12 ks to 75 ks, with a mean value of 43 ks. The nominal energy range (0.3-10 keV) is larger than Chandra and gives us the opportunity to better investigate both the soft X-ray region ($<$ 0.5 keV) and the properties of the K$\alpha$ iron line at 6.4 keV. \\All the \xmm{} observations has the simultaneous optical-UV photometry from Optical Monitor (\om{}, \citealt{om}). Four filters were used: U ($\lambda_{eff}=$ 344 nm), UVW1 ($\lambda_{eff}=$ 291 nm), UVM2 ($\lambda_{eff}=$ 231 nm) and UVW2 ($\lambda_{eff}=$ 212 nm). The simultaneous optical-to-X-ray coverage is a key information in the understanding of the interplay between the disk and its surrounding environment (BLR and corona).\\The data reduction was carried out with tools from SAS \citep{sas04,sas} v18.0.0 package. After the reprocessing\footnote{The calibration files were downloaded from the \xmm{} Science Archive.}, we filtered the flaring background, i.e., the detection of high-energy ($>$ 10 keV) charged particles that will cause a false-positive cascade. This filtering was performed by extracting the light curve of the events with energy above 10 keV and considering only the periods where the count rate is low ($\leq$ 0.4 counts/s, according to the \xmm{} users handbook\footnote{http://xmm-tools.cosmos.esa.int/external/xmm\_user\_support/\\documentation/uhb/XMM\_UHB.pdf}) and checked against these data. The removal of the flaring background caused the loss of the 46\% of the exposure time, reduced from 172.04 ks to 94.40 ks.\\The level of the pileup was evaluated by the task \textit{epatplot} and graded as negligible.\\The 0.3-10 keV source spectrum was taken from a 40'' wide circle equal to the 90\% and the 85\% of the MOS and pn cameras EEF at 5 keV, respectively. The background spectrum was extracted from a single source-free region. Data were grouped with a minimum of 25 counts per energy bin.

\begin{figure*}
	\resizebox{\hsize}{!}
    {\includegraphics[]{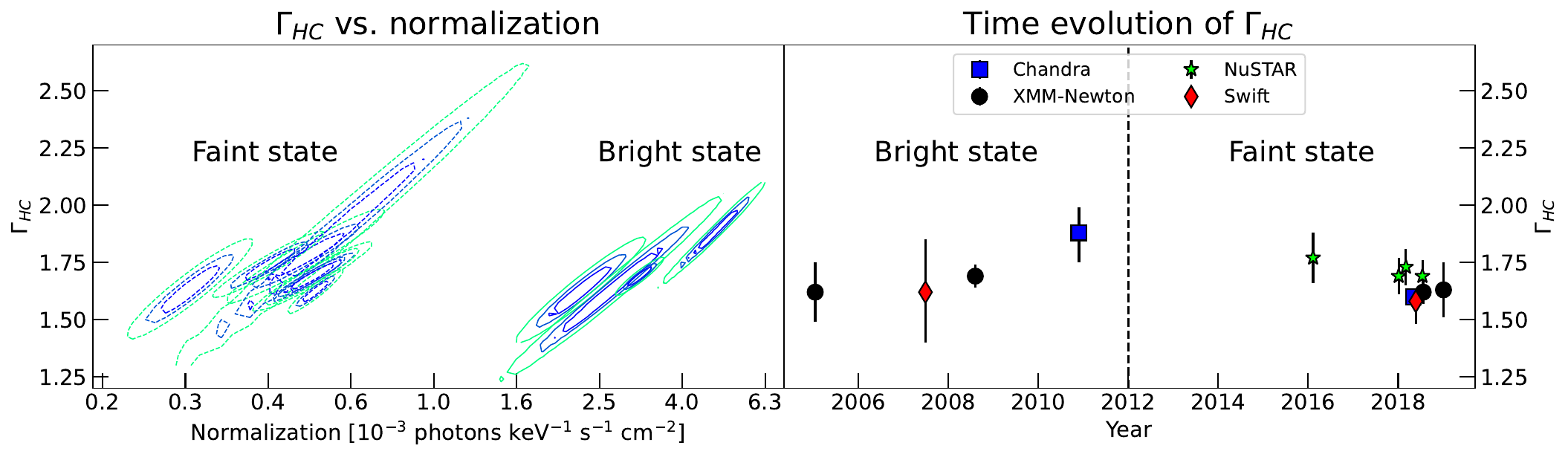}}
    \caption{Hot Comptonization properties of Mrk 1018. \textit{Left panel}: Confidence intervals of the photon index ($\Gamma_{HC}$) vs. the normalization for the hot Comptonization power law. Contour levels corresponds to confidence of 68\%, 90\% and 99\%. Dashed lines are used for the data related to the faint state, while solid lines correspond to the observations acquired during the bright state. \textit{Right panel}: time evolution of $\Gamma_{HC}$. Symbols and colours correspond to different telescopes: blue square \ch{}, black circle \xmm{}, green star \nus{} and red diamond \xr{}. The first data point of \ch{} corresponds to the observation 12868, which is heavily affected by the pileup. The vertical dashed black line denotes roughly the separation between the bright and the faint state.}
    \label{fig:gammanorm}
\end{figure*}

\subsection{\nus{} data reduction}
\nus{} \citep{nustar} acquired data of Mrk 1018 four times, one in 2016 and three in 2018, with exposure times that vary from 22 ks to 43 ks, with a mean value of 33 ks. Thanks to it nominal bandpass of 3-79 keV, we could investigate the high-energy region of Mrk 1018 spectrum.\\Because \nus{} use CZT (Cadmium-Zinc-Telluride) detectors, its data are always free from significant pileup. The spectra were extracted with the tools \textit{nupipeline} from the HEASOFT\footnote{http://heasarc.gsfc.nasa.gov/ftools} \citep{heasoft} v6.26.1 package, that produced filtered and calibrated\footnote{The calibration parameters were automatically retrieved from the online HEASARC archive CALDB.} events files, and \textit{nuproduct}, that built the image in the nominal 3-79 keV range and extract the source and background spectrum.\\The selection region for the source was a 40'' wide circle (50\% of the EEF at 10 keV for both cameras), while the background was defined by four circular regions free of sources around the image (angular distances and radii were arbitrary). The data were grouped with a minimum of 30 counts per energy bin.\\Inspecting the signal-to-noise ratio of the spectra, we found that the channels above 25 keV were extremely noisy and, consequently, we removed them losing the information about the very high-energy environment.

\subsection{\xr{} data reduction}
\xr{} (0.3-10 keV) has the largest set of observations, with a total of sixty-three data acquisition spanning the 2005-2019 period. However, it has the lowest exposure per observation. Indeed, the mean exposure time is 4 ks with minimum and maximum value of 0.7 ks and 7 ks, respectively. The time-wide coverage of Swift helped us to place constraints on the time-evolution of Mrk 1018 spectrum and flux.\\Also, we exploited the simultaneous \uv{} photometry obtained in six optical/UV filters: V ($\lambda_{eff}=$ 547 nm), B ($\lambda_{eff}=$ 439 nm), U ($\lambda_{eff}=$ 347 nm), UVW1 ($\lambda_{eff}=$ 260 nm), UVM2 ($\lambda_{eff}=$ 225 nm) and UVW2 ($\lambda_{eff}=$ 193 nm).\\Using the online tool `Swift Build XRT products routine'\footnote{https://www.swift.ac.uk/user\_objects/}, we combined together the observations from 2005 to 2010 to produce a dataset for the bright state, and from 2016 to 2019 resulting in a dataset for the faint state. The data were grouped with a minimum of 30 counts per bin.\\Observations 00049654001 and 00049654002 were taken in 2014 when the source was transitioning to the faint state. Consequently, given the very low statistics in their spectra we opted to not use them in the X-ray analysis.

\subsection{Optical/UV data reduction}\label{sec:uvred}
As far as the optical/UV data reduction is concerned, \om{} data were reduced by running the SAS task \textit{omichain} with default input parameters. We also extracted for each filter in each observation the spectral points using \textit{om2pha}.\\\uv{} photometry, instead, was retrieved with the task \textit{uvotsource} from the HEASOFT package using as input:
\begin{itemize}
    \item a source region equal to a 5$''$ wide circle (i.e., the point spread function size) centred on the source coordinates \citep{gaiadr3}. Those were retrieved from SIMBAD Astronomical Database \citep{simbad}, as suggested by \citet{pal16};
    \item a background region of $\sim$ 50$''$ away from the source;
    \item a 3$\sigma$ background threshold, where $\sigma$ is the root mean square noise of each image.
\end{itemize}
We also combined together with the task \textit{uvotimsum} the observations from 2005 to 2010 to obtain an image in each filter for the bright state, and from 2016 to 2019 to produce the image for the faint state. We finally run \textit{uvotsource} with the same input as above to get the photometry, and \textit{uvot2pha} to extract the spectral points, for each filter, for both the bright and faint state.

\begin{figure*}
	\resizebox{\hsize}{!}
    {\includegraphics[]{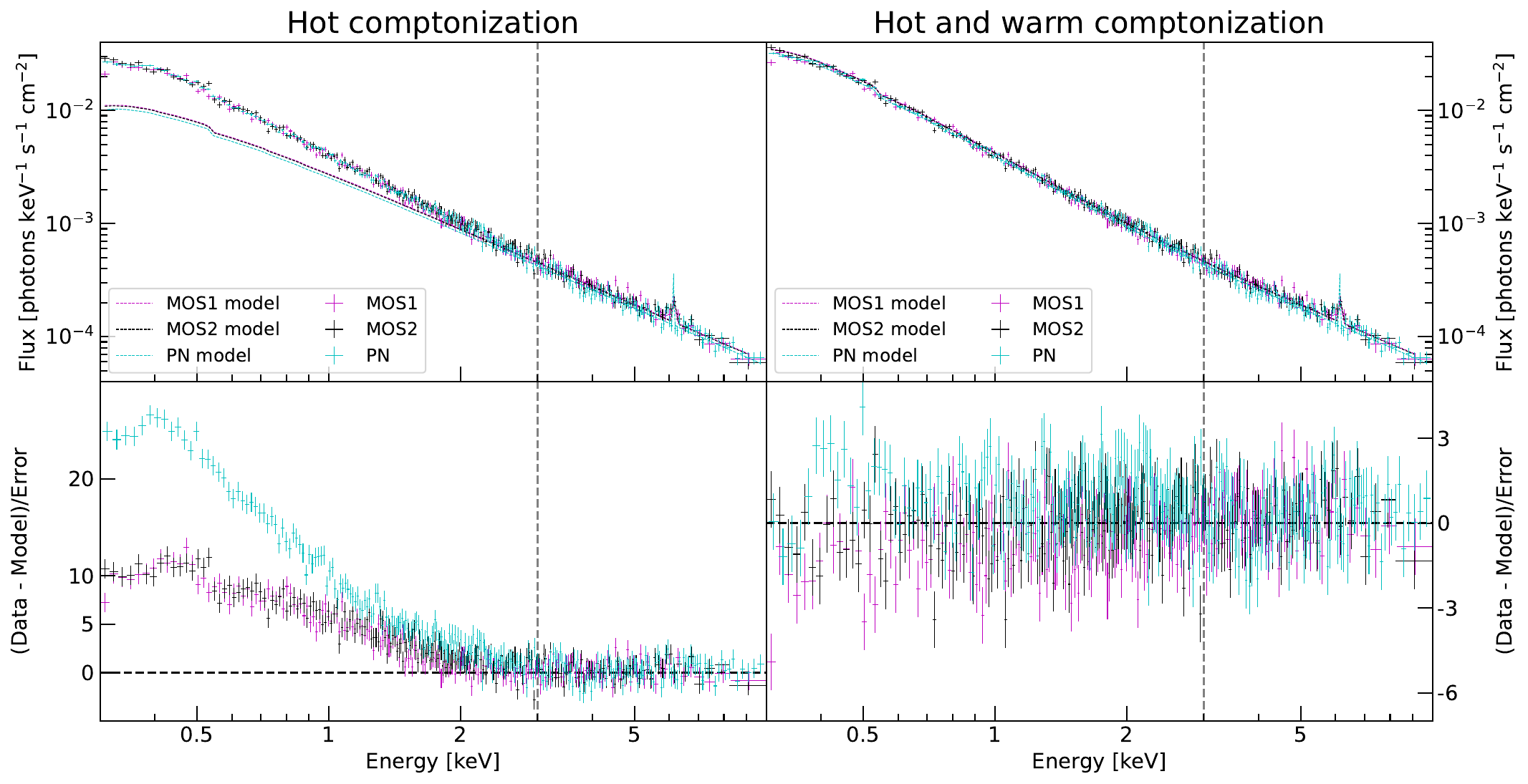}}
    \caption{Mrk 1018 0.3-10 keV spectrum taken from XMM 2008 observation (ID: 0554920301). \textit{Top left panel}: Spectrum and the best-fit model (\ref{eq:mo}) for each camera. The data of the two MOS are plotted in magenta and in black, while pn data are represented in dark cyan. The best-fit is reported using the bold lines (same colour as before) and is referred to data at energies above 3 keV (the vertical dashed grey line). \textit{Bottom left panel}: residuals, expressed in terms of $(data-model)/error$, for the hot Comptonization model (\ref{eq:mo}). The horizontal dashed black line indicates the 0-level, while the vertical dashed grey line denotes the 3 keV energy channel. \textit{Top right panel} and \textit{bottom right panel}: same as before, but now the model and the residuals are referring to the modelling with both the hot and warm Comptonization and fitted over the whole energy range.}
    \label{fig:spec}
\end{figure*}

\section{Data analysis}\label{sec:ana}
\subsection{X-ray spectroscopy}\label{sec:xspec}
\begin{figure}
	\centering
	\includegraphics[width=\hsize]{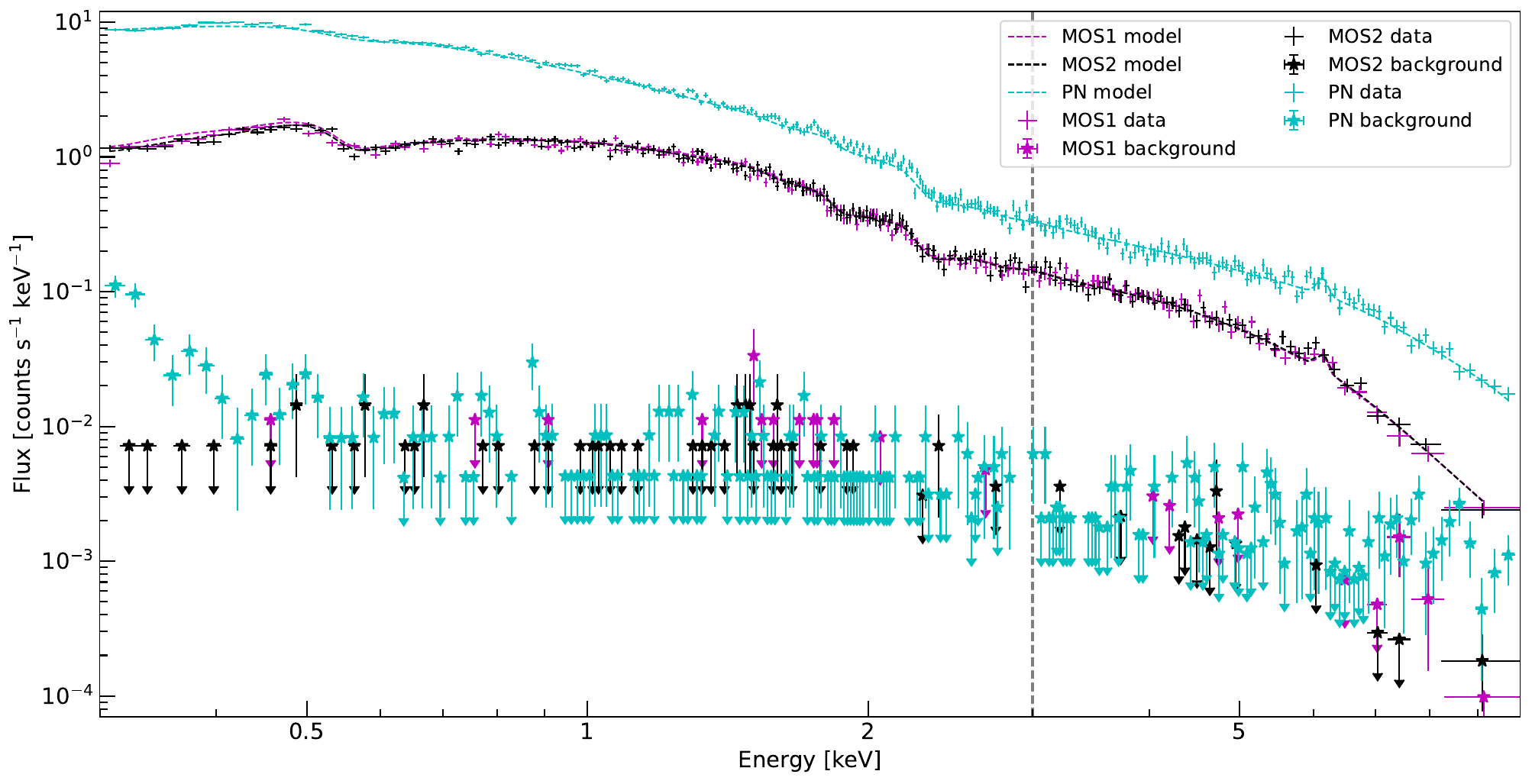}
    \caption{Comparison between Mrk 1018 0.3-10 keV spectrum and background level for the XMM 2008 observation (ID: 0554920301). The color-coding is the same as in Fig. \ref{fig:spec}. The background is negligible across the whole energy band for all the cameras.}
    \label{fig:back}
\end{figure}

\begin{figure*}
	\resizebox{\hsize}{!}
	{\includegraphics[]{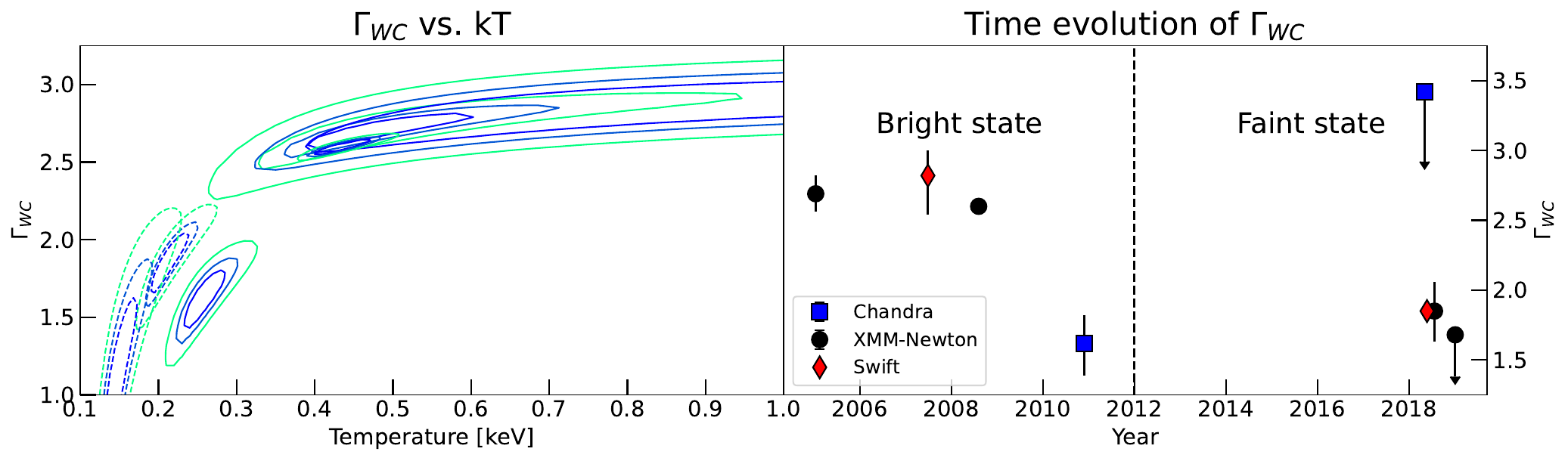}}
	\caption{Warm Comptonization properties of Mrk 1018. \textit{Left panel}: Confidence intervals of the warm Comptonization spectral index ($\Gamma_{WC}$) vs. the warm corona temperature. Contour levels corresponds to confidence of 68\%, 90\% and 99\%. Dashed lines are used for the data related to the faint state, while solid lines correspond to the observations acquired during the bright state. \textit{Right panel}: time evolution of $\Gamma_{WC}$. Symbols and colours correspond to different telescopes: blue square \ch{}, black circle \xmm{}, green star \nus{} and red diamond \xr{}. The first data point of \ch{} correspond to the observation 12868, which is heavily affected by the pileup. The vertical dashed black line denotes the separation between the bright and the faint state.}
	\label{fig:gammakt}
\end{figure*}

We carried out the spectral analysis using XSPEC v12.13.0c \citep{arnaud96}. We started the analysis of the X-ray spectra using a simple model with three components:
\begin{equation}\label{eq:mo}
model=phabs\cdot(zpw+zgauss)
\end{equation}
where:
\begin{enumerate}
    \item \textit{phabs}  is the effect of the photoelectric absorption, which depends on the column density ($n_H$) and the photoelectric cross-section. It is associated with the Milky Way gas; we imposed $n_H=$ 2.48 $\times$ 10\textsuperscript{20} atoms cm\textsuperscript{-2}, as given by the neutral hydrogen map of \citet{hi4pi};
    \item \textit{zpowerlw} is the power law originated from the inverse Compton scattering of the photons emitted by the accretion disk. We will refer to it as the `primary power law'. The free parameters are the spectral index ($\Gamma_{HC}$) and the power law normalization ($k_{HC}$).
    \item \textit{zgauss} models the 6.4 keV Fe K$\alpha$ emission line  using a Gaussian profile. We fixed the line width to 0.01 keV and let the normalization and the rest-frame energy free to vary.
\end{enumerate}
We fit the model only to energies 3-10 keV (3-25 keV for \nus{}), as in this regime the primary power law is likely dominant, and produced the confidence contours to study the variation of $\Gamma_{HC}$ as a function of the normalization (see Fig. \ref{fig:gammanorm}). The best-fits are acceptable, as the $\chi^2_{red}=\chi^2/\text{DOF}$ (Degrees of Freedom) is about 1.0 for each observation.\\When we apply these best-fits to the full energy range, however, we found evident positive residuals below 2 keV. As an example, we report in the left panels of Fig. \ref{fig:spec} the best-fit for the \xmm{} 2008 observation 0554920301. We modelled this soft excess first by using a thermal component (\textit{mekal}: \citealt{mewe85,mewe86,liedahl95}) representing the emission from the diffuse hot gas in the host galaxy, then via a double power law. However, the outcomes were in both cases not reasonable. Indeed, in the former the gas temperature was variable between the two states, which is not physically possible given the considered timescale of few years; in the latter the second power law was negligible.\\Therefore, we modelled the soft excess using the dual-corona model \citep{czerny87,mehdipour11,petrucci13,ursini20,middei20,alston23,kawanaka23}. In this scenario, there are a hot ($\sim$ 100 keV) optically thin ($\tau<1$) spherical corona around the SMBH producing the primary power law, and a warm ($<1$ keV) optically thick ($\tau\gg1$) corona in the innermost regions of the accretion flow that up-scatter the UV photons of the accretion disk to the soft X-ray range. However, the geometry of this secondary corona is still unknown \citep{done12,petrucci13}.\\We used the XSPEC model \textit{nthcomp} \citep{zdziarski96,zycki99} to reproduce the continuum of the warm comptonization (WC). It is described by five parameters: the power law spectral index\footnote{This parameter has not the same physical meaning of $\Gamma_{HC}$. It is a mathematical representation that quantifies the steepness of the soft excess bump at energies $<2$ keV.} ($\Gamma_{WC}$), the temperature of the seeding photons, the temperature of the corona and the normalization, i.e., the flux at 1 keV. We set the seeding photons, whose temperature was fixed at 22 eV given the SMBH mass of about $10^{7.84}\rm{ M}_{\odot}$, to be coming from a geometrically thin accretion disk emitting like a black body \citep{shakura73}.\\Thus, the observed spectra were described with the following model:
\begin{equation}\label{eq:mofinal}
    model=phabs\cdot(nthcomp+zpowerlw+zgauss)
\end{equation}
For each dataset we fixed the parameters of \textit{zpowerlw} and \textit{zgauss} to their best-fit value obtained before. For the \xr{} combined dataset of the faint state, the properties of the WC were not constrained, hence, we fixed the spectral index and the corona temperature to the values of the closest in time \xmm{} observation. The results are reported in Table \ref{table:result} and the improvement in the fits is shown in the right panels of Fig. \ref{fig:spec}.\\For completeness, we also checked the background spectra, which show that the noise level is negligible over the whole energy range. Again, as an example we present in Fig. \ref{fig:back} the one of \xmm{} 0554920301 dataset. This increases further the robustness of our analysis.\\We produced the confidence contours for the warm corona temperature against the WC spectral index (left panel of Fig. \ref{fig:gammakt}), and found that the datasets for each state are very isolated in the parameter space, with the exception of Chandra 12868 observation. This is the one heavily affected by the pileup, therefore the corresponding spectrum must be interpreted with caution (see Sec. \ref{sec:wcdisc} for further discussion).


\subsection{Optical, UV and X-ray photometry}\label{sec:photo}
\begin{figure}
	\centering
	\includegraphics[width=\hsize]{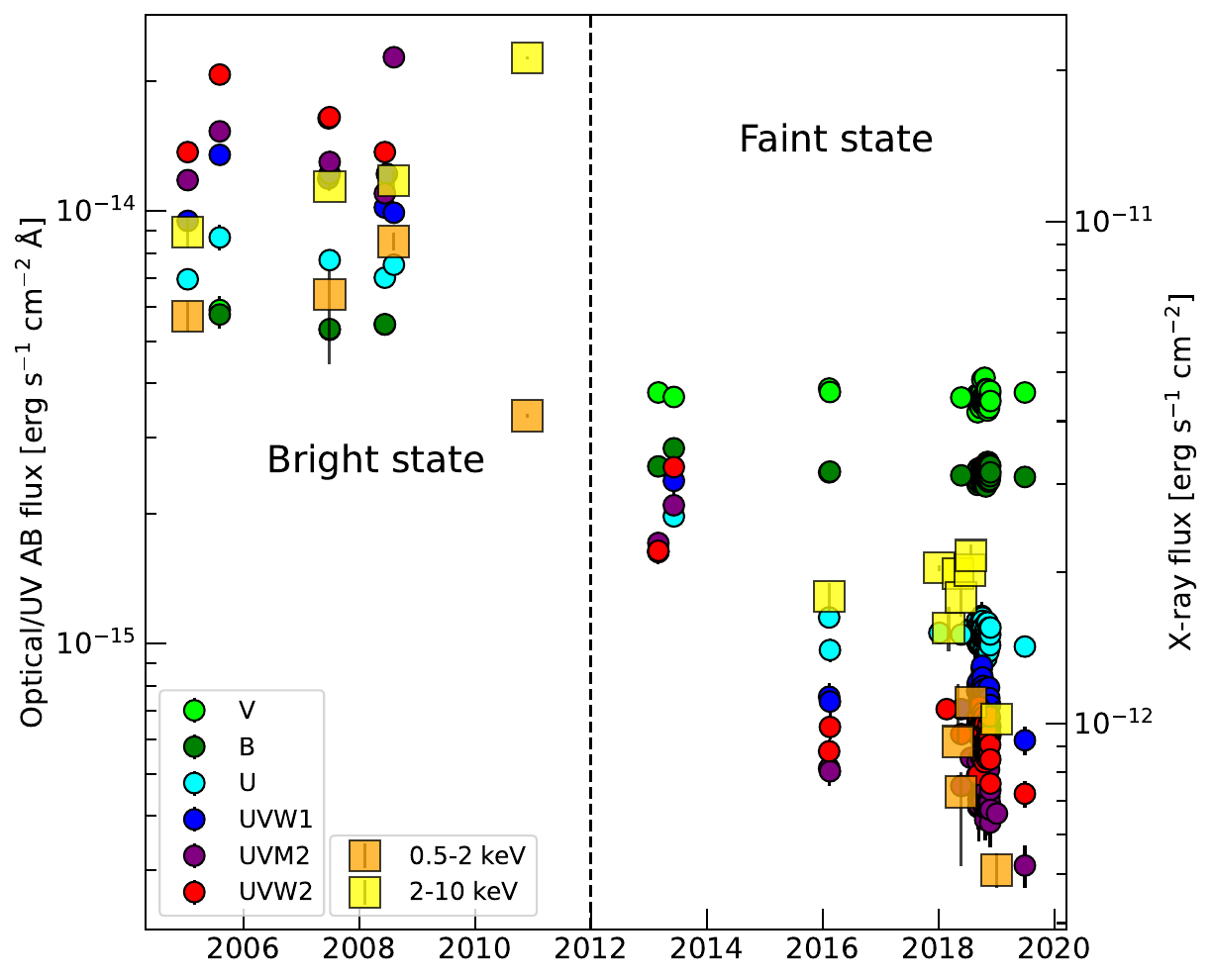}
      \caption{Optical, UV and X-ray evolution of Mrk 1018 photometry from 2005 to 2019. Values are reported in Tables \ref{table:obs} and \ref{table:om2}. The scales are different: the y-scale on the left is related to the optical/UV fluxes, the one on the right to the X-ray emission. The X-ray points are labelled with orange and yellow squares for the 0.5 - 2 keV and 2 - 10 keV band, respectively. Each optical/UV filter is marked with circles of different colours: $V$ in green, $B$ in dark green, $U$ in cyan, $UVW1$ in blue, $UVM2$ in purple and $UVW2$ in red. The vertical black dashed line distinguishes between bright and faint state.}
    \label{fig:flux}
\end{figure}

\begin{figure}
	\centering
	\includegraphics[width=\hsize]{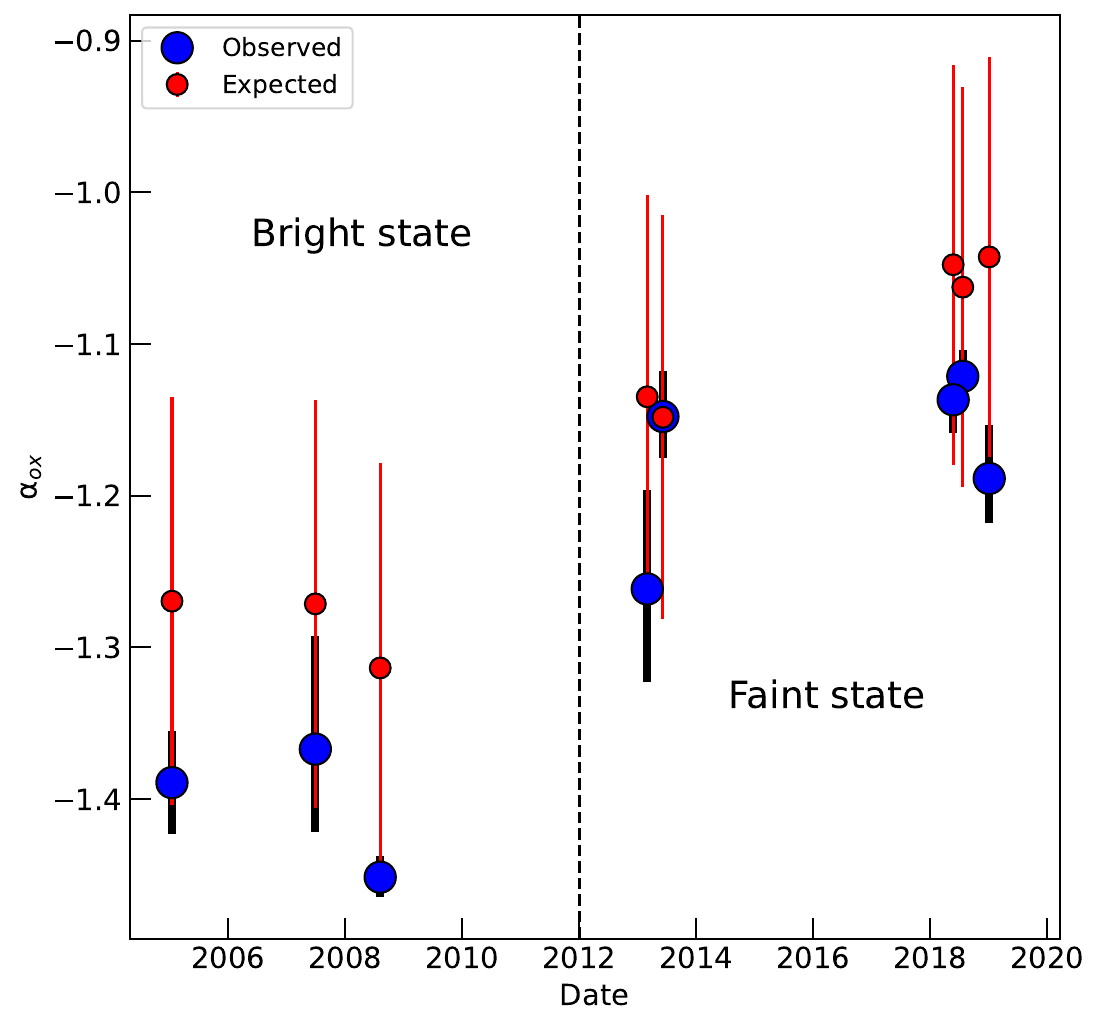}
	\caption{Comparison between Mrk 1018 observed (blue) and expected (red) spectral index $\alpha_{ox}$ over the 14 years of observations presented in this paper. The vertical black dashed line denotes the separation between the bright and the faint state. The uncertainties of the expected value of $\alpha_{ox}$ were calculated applying the propagation of uncertainties formula on equation \ref{eq:theo}.}
    \label{fig:aox}
\end{figure}

We used the optical and UV fluxes provided by the data reduction process of \om{} and \uv{} observations and the 0.5-10 keV fluxes retrieved from the best-fits of the X-ray spectra to study the photometric properties of Mrk 1018 between the two states. By plotting all the measurements (see Fig. \ref{fig:flux}), it is evident that the brightness of Mrk 1018 exhibits a variability, sometimes dramatic, in all bands, as reported in Tables \ref{table:obs} and \ref{table:om2} as well as summarized in Table \ref{table:ratios}.\\We checked if the UV and X-ray flux variations might be correlated, as we would expect also from the well-known AGN relation between the luminosity at 2500 \AA{} and at 2 keV (e.g., \citealt{zamorani81,strateva05,lusso10}). The interplay between UV and X-ray is parameterized by the spectral index $\alpha_{ox}$ defined as\footnote{$\alpha_{ox}$ is defined negative using equation (\ref{eq:aox}).}
\begin{equation}\label{eq:aox}
\alpha_{ox}=\frac{\log(F_{2\text{ keV}})-\log(F_{2500})}{\log(\nu_{2\text{ keV}})-\log(\nu_{2500})}=-0.384\log\Big{(}\frac{L_{2\text{ keV}}}{L_{2500}}\Big{)}
\end{equation}
To compute $\alpha_{ox}$, we took advantage of \uv{} UVW1 filter since it is the one with the closest effective wavelength ($\nu_{\lambda_{eff}}$) to 2500 \AA{}. $F_{2500}$ was estimated under the assumption that $F_\nu\propto\nu^{-0.5}$ \citep{vandenberk01}:
\begin{equation}\label{eq:fluxconv}
    F_{2500}=F_{\lambda_{eff}}\times\Big{(}\frac{\nu_{2500}}{\nu_{\lambda_{eff}}}\Big{)}^{-0.5}\text{ erg}\text{ cm}^{-2}\text{ s}^{-1}\text{ Hz}^{-1}
\end{equation}
Instead, $F_{2\text{ }keV}$ was extrapolated from the primary power law normalization, i.e. the flux at 1 keV, using:
\begin{equation}
    F_{2\text{ keV}}=F_{1\text{ keV}}\times2^{-\Gamma}\text{ erg}\text{ cm}^{-2}\text{ s}^{-1}\text{ Hz}^{-1}
\end{equation}
The time evolution of $\alpha_{ox}$ shows a slight increase in the value between the bright and faint state, going from $-$1.38 $\pm$ 0.02 to $-$1.18 $\pm$ 0.04 (see also Fig. \ref{fig:aox}). It means that the UV/X-ray ratio becomes smaller. We also note that the variation of $\alpha_{ox}$ were consistent with the relation, reported with the red points in Fig. \ref{fig:aox}, found by \citet{nanni17} between $\alpha_{ox}$ and $L_{2500}$:
\begin{equation}\label{eq:theo}
\alpha_{ox}=-(0.155\pm0.003)\log(L_{2500})+(3.206\pm0.103)
\end{equation}
This supports the hypothesis that the X-ray emission is responding to a reduction in the UV. Furthermore, this result allows us to speculate that something is happening in Mrk 1018 at the level of the accretion disk, since the UV photons come mainly from that region.\\As it will be discussed in the next sections, by combining our large amount of data with models reported in the literature, we will robustly conclude not only that the changing-look behaviour of Mrk 1018 is indeed likely due to a change of state of the inner accretion flow, but we will push our interpretation even further and reveal, for the first time, the complexity of mechanisms leading to such variations.

\section{Discussion}\label{sec:disc}
\begin{table*}
    \caption{Broad-band variability of Mrk 1018.}
    \label{table:ratios}
    \centering
    \begin{tabular}{c c c c c}
    \hline\hline
    Energy band & $\lambda_{eff}$ & Bright state flux & Faint state flux & Decrease factor\\
    V\tablefootmark{a} & 547 & 5.7 $\pm$ 0.2 & 3.7 $\pm$ 0.1 & 1.5\\
    B\tablefootmark{a} & 439 & 5.6 $\pm$ 0.1 & 2.5 $\pm$ 0.1 & 2\\
    U\tablefootmark{a} & 347 & 7.9 $\pm$ 0.8 & 1.1 $\pm$ 0.1 & 7\\
    UVW1\tablefootmark{a} & 260 & 12 $\pm$ 1 & 0.8 $\pm$ 0.2 & 15\\
    UVM2\tablefootmark{b} & 231 & 17 $\pm$ 5 & 0.5 $\pm$ 0.1 & 34\\
    UVM2\tablefootmark{a} & 225 & 13 $\pm$ 2 & 0.5 $\pm$ 0.3 & 26\\
    UVW2\tablefootmark{a} & 193 & 17 $\pm$ 3 & 0.7 $\pm$ 0.3 & 24\\        
    0.5-2 keV & - & 7 $\pm$ 2 & 0.8 $\pm$ 0.2 & 9 \\
    2-10 keV & - & 14 $\pm$ 4 & 1.8 $\pm$ 0.3 & 8
    \end{tabular}
    \tablefoot{Optical/UV fluxes are given in units of 10\textsuperscript{-15} erg s\textsuperscript{-1} cm\textsuperscript{-2} \AA\textsuperscript{-1}, while X-ray values are in units of 10\textsuperscript{-12} erg s\textsuperscript{-1} cm\textsuperscript{-2} and $\lambda_{eff}$ is expressed in nm. The reported numbers are, for each AGN state, the mean of the photometric measurements given in Tables \ref{table:obs} and \ref{table:om2}, while the errors correspond to the standard deviation.\\
        \tablefoottext{a}{\uv{}.}
        \tablefoottext{b}{\om{}.}
    }
\end{table*}

\subsection{The hot corona}
Fig. \ref{fig:gammanorm} shows the temporal evolution of $\Gamma_{HC}$ with respect to both the normalization of the HC component (left panel) and time (right panel). Over the course of the analysed 14-years period, the HC spectral index has remained nearly constant, indicating minimal variations in the interplay between the hot corona and the accretion disk \citep{haardt91,alston23}.\\However, there is a substantial variation in the normalization of the power law, with a decrease of a factor $\sim10$. This dimming of the coronal emission is likely caused by the fainter flux of UV seed photons, as the analysis of $\alpha_{ox}$ is suggesting, in turn probably related to a lower accretion rate rather than a significant increase in the line-of-sight column density.\\Indeed, adding an absorption term to model (\ref{eq:mofinal}) resulted in a negligible obscuration. A recent investigation of the ultra-hard X-ray light curve using \textit{Swift}-BAT \citep{temple23} has demonstrated a gradual decline (from $\sim1$ mCrab to $<0.1$ mCrab) in the 14-195 keV flux over a 13-year timescale, unrelated to an increasing column density along the line of sight. Other studies have also achieved the conclusion that an eclipsing event is likely not the origin of the changing-look behaviour in Mrk 1018 \citep{mcelroy16,husemann16,lamassa17,noda18,hutsem20,brogan23}.

\subsection{The warm corona}\label{sec:wcdisc}
Fig. \ref{fig:gammakt} illustrates notable differences in $\Gamma_{WC}$ between the bright ($\Gamma_{WC}\sim2.6$) and faint state ($\Gamma_{WC}\sim1.6$). The flattening of the WC power law indicates that the soft X-ray emission (E $<2$ keV) has dimmed more than the hard one during the state-transition. In other words, the X-ray spectrum became harder, as confirmed by the evolution of the hardness ratio
\begin{equation}\label{eq:h}
    H=\frac{F_{2-10\text{ keV}}-F_{0.5-2\text{ keV}}}{F_{2-10\text{ keV}}+F_{0.5-2\text{ keV}}}
\end{equation}
that went from 0.2 $\pm$ 0.1 during the bright state to 0.4 $\pm$ 0.1. We will return to this in Sec. \ref{sec:hypothesis}.\\Fig. \ref{fig:gammakt} also reveals important variations in the warm corona temperature, which decreases from $\sim0.4$ keV to $\sim0.2$ keV. These changes strongly suggest a potential connection between the changing-look behaviour of Mrk 1018 and the properties of the warm comptonizing medium.\\Intriguingly, \ch{} observation 12868 captured Mrk 1018 in the bright state, yet the warm corona temperature is similar to those observed during the faint state (see Table \ref{table:result}). This might imply that the cooling of the warm corona occurred within a time interval of less than two years, as the spectrum of the 2008 \xmm{} observation (obsID: 0554920301) was adequately described by a warm corona with a temperature of $\sim0.4$ keV.\\The results presented so far rise the following question: what is the physical process capable of cooling the warm corona?

\subsection{The role of magnetic fields}\label{sec:bfield}
Previous studies have proposed that the physical mechanism governing the corona temperature is likely correlated with magnetic fields \citep{merloni01,reeves02,fabian15}. Thus, the observed cooling of the warm corona between the two states may be a consequence of the weakening in the magnetic intensity.\\In this scenario (see \citealt{ferreira03} for the mathematical derivation, and \citealt{ferreira06,marcel18a,marcel18b,marcel19,marcel20} for its applications to stellar-mass black hole X-ray binaries),  the lines of the magnetic field in the accretion disk are oriented vertically with respect to the disk plane. As the disk is rotating, a torque is applied to the lines, making them bend horizontally. In turn, this bending generates another torque acting radially, bringing mass inward. Hence, the viscous timescale is effectively being reduced, and the stronger the magnetic field, the stronger the radial torque and the lower the timescale for accretion.\\Applying this model to Mrk 1018 results not only in a reduced intensity of the magnetic field between the bright and the faint state, but also in a natural explanation of the very short timescale ($\sim10$ years) for the variability \citep{feng21}. Therefore, this discussion strongly points toward the magnetic fields as responsible for the photometric variations observed for this source: as they weaken, the radial torque become less significant, and the radial inflow diminish. The result is a decrease in the UV luminosity emitted by the AGN, which strongly correlates with the accretion rate. The dimming in UV luminosity, in turn, reduces the number of seed photons available for the HC, leading to a fainter X-ray emission. By invoking the magnetic field we are able to explain with a single physical mechanism the broad-band photometric variability of Mrk 1018.

\subsection{The changing-state accretion disk}\label{sec:hypothesis}
There are still two open questions that we need to answer. First, what is causing the spectral variability, and second, what is lying behind the weakening of the magnetic fields. Both questions might be answered by arguing that the changing-look behaviour of Mrk 1018 is due to a changing-state accretion disk. This idea has already been proposed, for example, by \citet{noda18} and \citet{dexter19a} for this source, and by \citet{stern18} for the CL-AGN WISE J105203.55+151929.5, and we think our analysis has revealed another element in favour of this interpretation.\\Indeed, the fact that during the bright state the X-ray hardness ratio of Mrk 1018 was lower than in the faint state (see Sec, \ref{sec:wcdisc}), remarkably resemble the behaviour of stellar-mass black hole X-ray binaries when they evolve from an active to a quiescent phase (e.g., \citealt{fender04,petrucci08,begelman14}). In this transition, the accretion disk changes its state from a geometrically thin, optically thick structure (Shakura-Sunyaev Disk, SSD, \citealt{shakura73}) to a geometrically thick, optically thin flow (Advective Dominated Accretion Flow, ADAF, \citealt{narayan96}). In an ADAF, the energy is primarily released through the formation of a jet \citep{rees82,narayan95,narayan96,yuan04,yuan09,taam12}, which removes magnetic energy from the system and effectively weakens the magnetic fields. We argue that this transition is occurring in the inner part of the accretion flow, because in the faint state the hard X-ray spectral shape is unchanged.\\Therefore, we are now able to explain both the spectral and the photometric variability of Mrk 1018: the transition of the accretion disk from a fully SSD to an outer SSD with an inner ADAF changes the mode of energy release in the inner accretion flow, which now occurs primarily in the form of a jet. The jet extracts magnetic energy from the system, reducing the intensity of the magnetic fields.\\From this, two main outcomes can be derived. First, the warm corona cools down, and the soft X-ray emission is no more fully supported by it, causing the 0.5-2 keV flux to drop. Second, the weaker magnetic fields lowers the accretion rate, leading to the broad-band dimming of the AGN as described at the end of Sec. \ref{sec:bfield}. The former is likely the origin of the observed X-ray spectral variability observed through the comparison of the hardness ratio.\\The changing-state accretion disk scenario might also explain the optical spectral variability, that is, the type 1-type 2 transition. \citet{mcelroy16} argued that the observed disappearing of the broad optical lines could be potentially related to the alteration of the BLR structure and kinematics due to a binary SMBH system. However, based on our results, we think that what is affecting the BLR properties is, instead, the presence of a jet. Nonetheless, even our conclusion in this regard needs further data to be confirmed.

\section{Confirming the disk state-transition}\label{sec:scenario}
To validate the hypothesis that the changing-look events observed in Mrk 1018 are attributed to the transition of the accretion disk from a fully radiative flow to a `quasi-radiative' disk with an inner advective region and vice versa, we performed a Spectral Energy Distribution (SED) fit. This kind of analysis has been already carried out by \citet{noda18}, but their results were limited by the small sample of data. Instead, we can rely on a much larger set of data points, both in terms of energy range (we have the $>$ 10 keV measurements provided by \nus{}, which were not available at their time of writing), temporal coverage (2005 to 2019 in comparison to their 2008 to 2016) and statistics.

\subsection{SED fitting}\label{sec:sed}
The SED fit incorporated optical/UV spectral points\footnote{The V, B, and U filters were not used extensively due to contamination from the host galaxy emission.} obtained from all \om{} observations and from the combined \uv{} images (see Sec. \ref{sec:uvred}), along with the majority of the X-ray data listed in Table \ref{table:obs}. We left out from the fitting \ch{} ObsID 12868, because the source might have been captured at the beginning of the state-transition (see the discussion in Sec. \ref{sec:wcdisc}), and \xr{} datasets 00049654001 and 00049654002 because were taken during the transition.\\We simultaneously fitted the SED of each state using the XSPEC model \textit{AGNSED} \citep{narayan95,kubota18}. To reduce degeneracy and facilitate the fitting procedure, certain parameters were fixed to standard values. These include setting the dimensionless black hole spin to 0, the cosine of the inclination angle for the WC to 0.5, the hot corona temperature to 100 keV, the outer radius of the disk to the self-gravity radius, the upper limit of the scale height for the hot corona to the default value of 10 gravitational radii ($R_g=\frac{GM}{c^2}$), and the normalization to 1 (internally computed).\\The free parameters are so the SMBH mass, the accretion rate normalized to the Eddington ratio $\mu=\frac{\dot{M}}{\dot{M}_{edd}}$, the warm corona temperature and radius, the hot corona radius, and the spectral indexes for the WC and HC power laws. Additionally, we incorporated the Fe K$\alpha$ transition line, whose normalization was left free to vary while the width and the rest-frame energy were fixed at 0.01 keV and 6.4 keV, respectively; a dereddening factor $E(B-V)=0.372$ \citep{noda18}; the photoelectric absorption from the Milky Way gas, and a constant for each X-ray spectrum. This last term accounts for intrinsic flux differences between observations taken at significantly different times. The best-fit values of the physical parameters of the model are reported in Table \ref{table:agnsed}, and a plot showing the SED in both states is provided in Fig. \ref{fig:sed}.

\subsection{The bright state: an SSD disk}
\begin{figure*}
	\resizebox{\hsize}{!}
    {\includegraphics[]{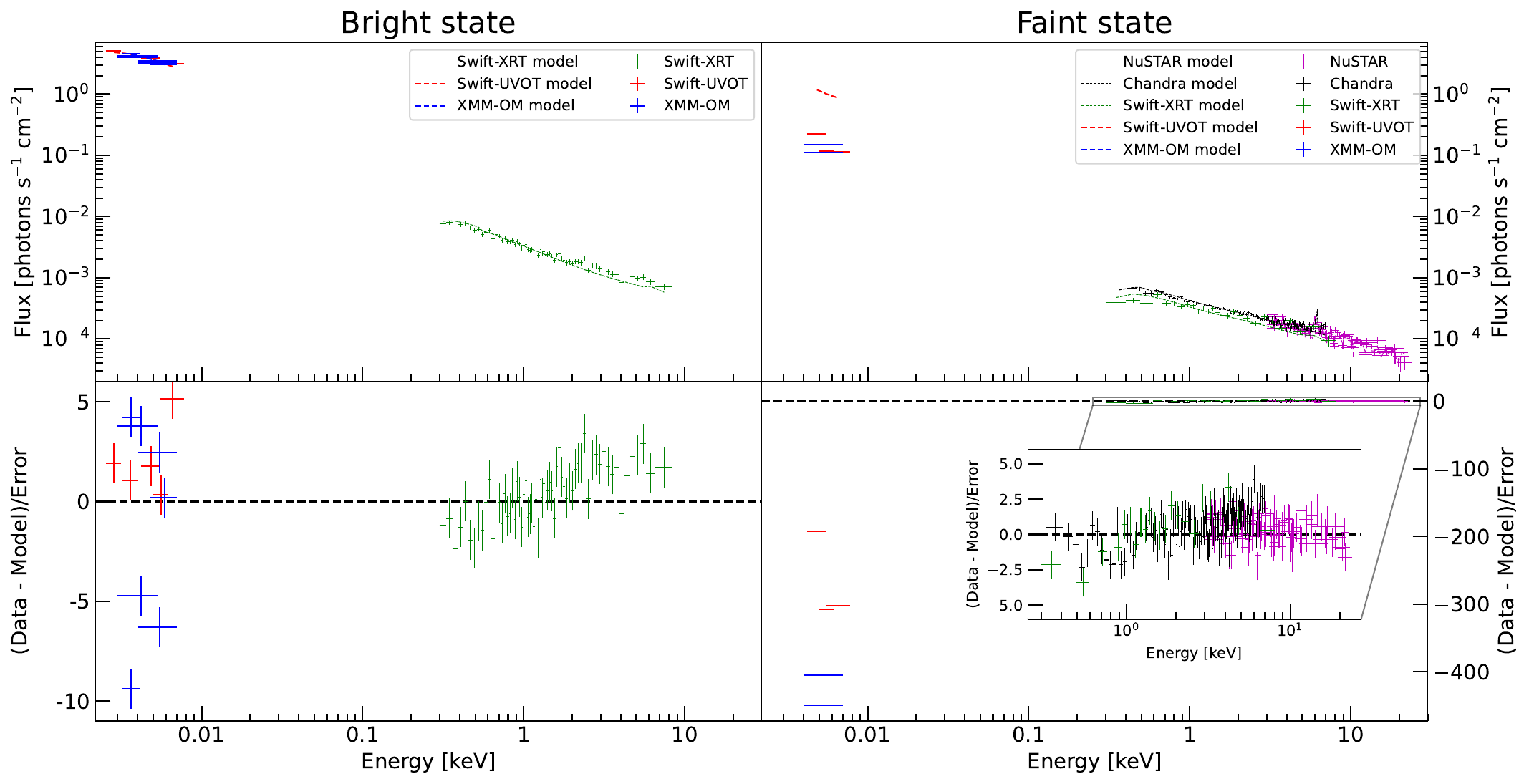}}
    \caption{Mrk 1018 optical-to-X-ray SED. \textit{Top left panel}: SED for the bright state. \xr{} data are reported in green, \uv{} in red and \om{} in blue. The best-fit model for each telescope is reported using the bold lines (same colour as before). \xmm{} data are not shown to avoid overcrowding. \textit{Bottom left panel}: residuals, expressed in terms of $(data-model)/error$, for the model described in Sec. \ref{sec:sed}. The horizontal dashed black line indicates the 0-level. \textit{Top right panel} and \textit{bottom right panel}: same as before, but now the data and residuals refers to the faint state. \nus{} points are given in magenta, while \ch{} in black. Y-axis scale is the same for the top panels, allowing for a direct comparison between the two states. The optical/UV points for the faint state are not fitted to the model, as the fit will otherwise being unconstrained. The residuals for the X-ray points are in the range $\pm5$, as shown in the inset panel.}
    \label{fig:sed}
\end{figure*}

The best-fit to the bright state of Mrk 1018 (left panels of Fig. \ref{fig:sed}) confirms that during that period around the SMBH there were an inner HC region extending up to $\sim20R_g$, a WC structure ($kT\sim0.46$ keV) with a size of $\sim65R_g$ and an SSD. The spectral index for the WC ($\Gamma_{WC}\sim2.64$) and the HC ($\Gamma_{HC}\sim1.71$), as well as the temperature of the WC region, align with the values obtained from our previous spectral analysis (refer to Sec. \ref{sec:xspec}), confirming the reliability of our best-fit model.\\The estimated SMBH mass of $\sim9.7\times10^{7}M_\odot$ is consistent with the estimates reported in the literature (see, for example, Table \ref{table:obj}). Additionally, we determined the Eddington ratio by calculating the bolometric luminosity at 3000\AA{}, as measured from the \uv{} UVW1 filter flux density:
\begin{equation}
    \mu=\frac{0.75\times5.18\times(4\pi D_{L}^2)\times3000F_{3000}}{1.26\times10^{38}M_{SMBH}}
\end{equation}
where 0.75 is the correction factor accounting for the viewing angle of the accretion disk \citep{runnoe12a}, 5.18 is the bolometric factor as retrieved from Table 1 of \citet{runnoe12a} and $D_L$ is the luminosity distance. The conversion from the effective wavelength of the UVW1 filter to 3000\AA{} was performed using Eq. (\ref{eq:fluxconv}). By plugging $D_L=183.5$ Mpc, we obtained $\mu=0.04\pm0.01$ for the bright state, closely aligned with our best-fit value.\\Based on the Eddington ratio, we conclude that during the bright state, the accretion disk in Mrk 1018 exhibited characteristics of a geometrically thin, optically thick accretion flow. This is supported by $\mu$ exceeding the critical value of $\mu_c=0.02$ \citep{kubota18}. Note that for the bright state $\mu\sim\mu_c$, promoting the idea suggested by \citet{noda18} that CL-AGN accrete close to $\mu_c$. Furthermore, the observed value of $\Gamma_{HC}\sim1.7$ indicates that the accretion disk in the bright state is likely truncated at the boundary of the inner hot corona (refer to Fig. 6 of \citealt{kubota18}), again living no space for an inner ADAF.


\subsection{The faint state: an SSD + ADAF disk}
We attempted to fit the SED of the faint state of Mrk 1018, with limited success. We observed that the optical/UV points in the SED were significantly overestimated, while the X-ray data points provided better constraints, although the $\chi^2_{red}$ was still $>>1$ for all the dataset. Consequently, we decided to apply our model exclusively to the X-ray spectra. This choice has no impact on the conclusions presented in Sec. \ref{sec:photo} about $\alpha_{ox}$. Indeed, $\alpha_{ox}$ is a measure of the correlation between the UV and X-ray flux, not an indicator of the underlying spectral shape between the 2500\AA{} and 2 keV flux. In other words, $\alpha_{ox}$ tells if the UV and X-ray emission are related, not how.\\As the fitting was still not converging, we fixed the SMBH mass and the hot corona properties to the best-fit values obtained for the bright state of Mrk 1018. This assumption is motivated by the physical reasoning that no variations occurred either in the SMBH or the hot corona between the two states. Our method yielded a good quality fit with a $\chi^2_{red}\sim1.06$.\\When we incorporated the optical/UV points into the best-fit model, they were once again found to be highly overestimated ($\chi^2_{red}>10^4$, see right panels of Fig. \ref{fig:sed}). This indicates that the optical/UV emission no longer originates from a radiative disk because it is considerably fainter than what is predicted by the model \textit{AGNSED}, which assumes a radiative disk. Indeed, both the Eddington ratio derived from the fit ($\mu=0.011\pm0.001$) and from the bolometric luminosity ($\mu=0.005\pm0.001$) are sub-critical, meaning the accretion flow is expected to be (partially) advective.\\If a portion of the flow is no longer efficiently cooling, the majority of the energy is released through a jet, as proposed by various studies \citep{rees82,narayan95,narayan96,yuan04,yuan09,taam12}. Consequently, the peak of emission from this medium shifts to longer wavelengths instead of the optical/UV band. As already anticipated, this state-transition is likely occurring in a small region of the accretion flow, probably the inner part. The results from the Chandra observation (ObsID: 12868) discussed in Sec. \ref{sec:wcdisc} could support this view. That observation is best described by an SSD but with a cooler WC region. As the WC medium is likely localized in the inner part of the accretion flow, the observation was probably taken when the inner ADAF was formed but the HC and the outer SSD did not have fully responded to that transition.\\Our in-depth analysis has finally managed to confirm the idea proposed by \citet{noda18} that the changing-look behaviour of Mrk 1018 is triggered by a state-transition of its (inner) accretion flow. In the following section, we are going to provide further elements that supports this conclusion.

\subsection{Elements in favour of disk state-transition}
A study conducted on Centaurus A using \nus{} and \xmm{} by \citet{furst16} concluded that the high-energy spectrum ($E>10$ keV) could be indicative of the accretion mode: a simple power law (or a single Comptonization component) suggests the presence of an inner ADAF. Indeed, if the accretion disk is entirely geometrically thin, we would expect to see the Compton hump at $E>10$ keV, due to the soft X-ray photons of the hot corona being reflected from the accretion disk \citep{magdziarz95}. Yet, if the inner accretion flow is puffed up, it can eclipse the hot corona and the reflection cannot take place anymore. The Compton hump disappears and the high-energy spectrum is described by only the Comptonization component. Our analysis of the faint state of Mrk 1018 indicates a single Comptonization component is sufficient to fit the \nus{} spectra, as shown in Table \ref{table:result}, and there is no need for a reflection factor. However, the lack of \nus{} observations during the bright state limits our ability to draw firm conclusions.\\Another element supporting the state-transition is provided by the polarimetry analysis of \citet{hutsem20}. They compare the polarimetric properties of the broad H$\alpha$ and $H\beta$ in the bright and faint state of the AGN, finding some intriguing results. First of all, the continuum polarization did not change significantly, and it was even slightly higher in the bright state, rejecting again the hypothesis of an increasing line of sight column density as the source of the AGN variability. Secondly, the polarization signatures of a binary SMBH system \citep{savic19a,savic19b} were not observed. Lastly, they speculated that the asymmetric polarization profile of $H\alpha$ and the lack of a rotation in the polarization angle may suggests that the broad $H\alpha$ is produced in a polar outflow. We think that this conclusion could support \citet{chen21} interpretation of the presence of magnetic outflows in the disk, originated from the inner accretion flow becoming an ADAF.\\\citet{brogan23} reported a remarkable outburst event in 2020, where the fluxes in the \textit{u'}-, \textit{c}- and \textit{o}-band exhibited an extraordinary increase by factors of $\sim$ 12, 22, and 14, respectively. They ruled out a number of possibilities to explain this outburst, from reduction in the line-of-sight absorption to a tidal disruption event (TDE, \citealt{rees88,phinney89}), a warped accretion disk \citep{rees88,phinney89} and a clumpy accretion flow \citep{natarajan98,raj21}. We think that this outburst might indicate the presence of an inner ADAF. Indeed, it is expected that at the interface between the outer SSD and the ADAF, radiation pressure instabilities arises, leading to quasi-periodic outbursts spanning a range of amplitudes and timescales \citep{sniegowska20}. A long-scale monitoring of Mrk 1018 during its faint state might potentially confirm this conclusion.\\Finally, recent Very Long Baseline Interferometry (VLBI) observations this AGN \citep{walsh23} have shown that the radio source associated with the nucleus has a proper motion inconsistent with a recoiling SMBH \citep{kim18} or a binary SMBH \citep{husemann16}, but likely with an unresolved jet component, as expected from the changing-state accretion disk scenario.

\begin{figure*}
	\resizebox{\hsize}{!}
    {\includegraphics[]{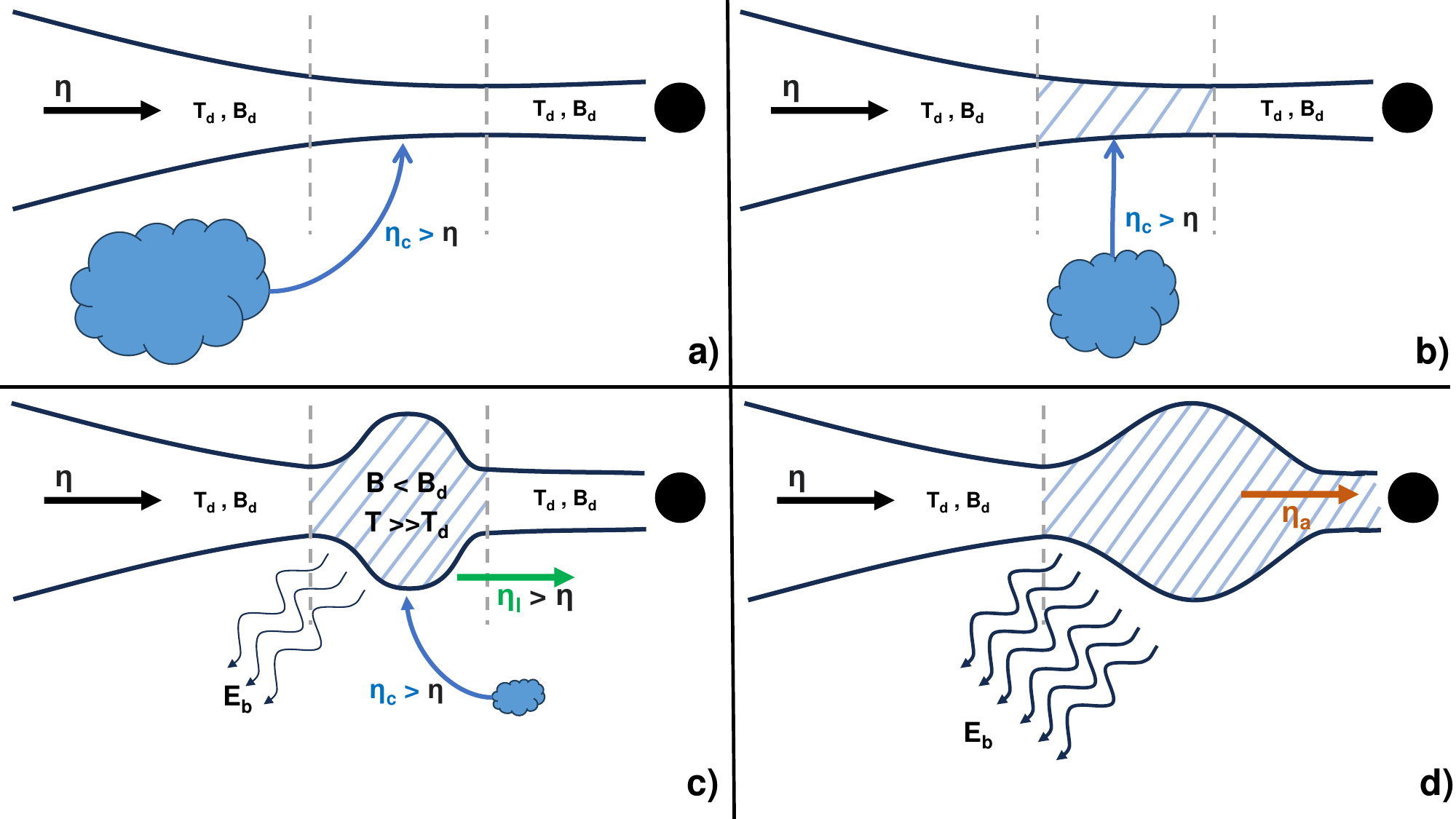}}
    \caption{Cartoon describing our proposed scenario to explain the broad-band spectro-photometric variability of Mrk 1018. The initial condition is an optically thick, geometrically thin disk having an internal accretion ($\eta$), temperature ($T_d$) and magnetic field ($B_d$). A cloud pushed toward the center, as an effect of a wet merger or the cold chaotic accretion, pass nearby and start to deposit gas at a given radius (denoted by the vertical dashed lines) with a rate $\eta_c$ (panel a). If $\eta_c>\eta$, the region in the disk where the gas from the cloud accretes starts to puff up. The energy produced via viscous friction is no more efficiently released and the temperature $T$ suddenly rises (panel b). When $T\gg T_d$, the hot gas is accreted inward at a rate $\eta_l>\eta$, triggering this instability in the adjacent inner ring. (panel c). The energy starts to be released magnetically and the magnetic field ($B$) in this region decreases. If $\eta_c\gg\eta$, this process extends to the whole inner accretion flow, making it optically thin and geometrically thick. The energy is released magnetically ($E_b$), likely in the form of a jet, drastically reducing the temperature of the gas and advecting the material onto the SMBH at a rate $\eta_a<\eta$ (panel d).}
    \label{fig:cartoon}
\end{figure*}

\subsection{The triggering mechanism of the state-transition}\label{sec:speculation}
One last thing we need to do is to provide a comprehensive, physically reasonable process which is able to trigger the changing state of Mrk 1018 accretion disk. The environment few parsecs around a SMBH is dense of gas, stars and stellar remnants \citep{ostriker83,syer91,artymowicz93,goodman04,levin07,nayakshin07,mckernan11a,mckernan11b}. Some of this material might enter the inner ($<$ pc) region where the accretion disk is located (e.g., \citealt{bellovary16,stern18}). This can happen in a number of different ways.\\For instance, the host galaxy of Mrk 1018 is likely a late-stage merger and in the case of a wet merger, gas is expected to lose angular momentum and feed the central SMBH \citep{volonteri05,hopkins06,hopkins08,kormendy13,gao20}. Another mechanism that can pull gas toward the accretion disk is the Cold Chaotic Accretion (CCA) \citep{gaspari13,gaspari17,gaspari20,maccagni21}. In this model, multi-phase clouds condense in and outside the galactic halo and descend onto the central SMBH. Within distances of $\sim$ 1 to 100 parsecs from the SMBH, these clouds undergo inelastic collisions, shedding angular momentum and plunging onto the SMBH. Furthermore, dynamical friction is also able to make massive objects like star clusters and gas clouds to lose angular momentum and sink onto the innermost regions of the AGN (e.g., \citealt{chandrasekhar43a,chandrasekhar43b,chandrasekhar43c,tremaine75,ostriker99}).\\We speculate that for Mrk 1018 what caused the state-transition was in fact probably a gas cloud. The reason is the duration of the bright state being too long to be caused by the disruption of a compact object (see, e.g., \citealt{mcelroy16}). We would like to note that a similar event, where a cloud passes close ($<$ pc) to a SMBH, is plausible, as it has been observed in our Galaxy \citep{valencia15}.\\In the following, we will try to explain how the close-by passage of a gas cloud can be the triggering mechanism for the state-transition of Mrk 1018 disk. Our hypothesis is schematized in the cartoon of Fig. \ref{fig:cartoon}. The close-by passage of such cloud establishes an accretion flow ($\eta_c$) onto the accretion disk, which in turn is accreting onto the SMBH with a rate $\eta$ (panel a). If $\eta_c>\eta$, the mass coming from the cloud will start to accumulate in this region, puffing it up, because the matter has no other options than extents vertically. The optical depth increases, and the energy produced through viscous friction is no more efficiently released (panel b). Therefore, the temperature in this region starts to increase, to the point that it induces a jump in the sound speed, enabling matter to move faster throughout the disk (see the discussion about H-ionization in \citealt{noda18}). As a result, in this region the local accretion rate ($\eta_l$) can become higher than $\eta$, triggering the instability of the subsequent innermost ring (panel c). If $\eta_c\gg\eta$, this local instability can lead to a switch in the properties of the whole inner accretion flow, which becomes an ADAF (panel d).\\This state-transition causes two outcomes (see discussion in Sec. \ref{sec:bfield} and Sec. \ref{sec:hypothesis}): a number of different instabilities \citep{noda18,stern18,dexter19a,sniegowska20,feng21}, which reduce the viscous timescale to the observed value of $\sim$ 10 years, and the formation of a jet \citep{rees82,narayan95,narayan96,yuan04,yuan09,taam12}, which weakens the magnetic fields and interacts with the surrounding environment, producing the broad-band spectro-photometric variability of Mrk 1018, including the type 1-type 2 evolution.\\We would like to stress that our speculative scenario is also qualitative. Factors that must be considered to confirm that a close-by passage of a gas cloud could trigger a state transition of the accretion disk include, for example, the mass of such a cloud, its dynamics with respect to the accretion disk and its density.  An optimal choice of these parameters is beyond the scope of this work and would require extensive simulations, that could be the subject of a follow-up study.\\We must also point out that the idea described above might be relevant only for AGN that are accreting very close to the critical Eddington ratio $\mu=0.02$ \citep{kubota18}. Indeed, for $\mu\gg0.02$ (radiative) and $\mu\ll0.02$ (jetted) AGN, the accretion disk is in equilibrium for its particular configuration (SSD for radiative AGN, SSD+ADAF or ADAF for jetted AGN) and any small perturbation will not produce any relevant variability in their emission. However, for AGN accreting close to $\mu=0.02$, even a tiny perturbation of the accretion rate might trigger a state-transition and a changing-look behaviour. This could also explain the rarity of CL-AGN.

\section{Conclusions}\label{sec:conc}
In this study, we conducted a comprehensive investigation of the changing-look AGN Mrk 1018, using an extensive dataset comprising optical, UV, and X-ray spectro-photometric measurements (Sec. \ref{sec:obs}). By performing a meticulous and systematic analysis of the X-ray spectra (Sec. \ref{sec:xspec}), broad-band photometry (Sec. \ref{sec:photo}), and optical-to-X-ray SED fitting (Sec. \ref{sec:sed}), we arrived at a compelling conclusion regarding the nature of not only Mrk 1018 variability, but possibly of the changing-look phenomenon in general.\\The changing-look behaviour is likely due to the inner region of the accretion flow undergoing a state-transition, becoming optically thin and radiatively inefficient, as described in our Sec. \ref{sec:hypothesis}. The formation of an inner ADAF leads to the internal energy being released through a jet, weakening the magnetic fields and establishing the spectro-photometric variability observed in Mrk 1018 (see Sec. \ref{sec:bfield} and Sec. \ref{sec:hypothesis}). In particular, we speculate that this AGN, and maybe most of the other CL-AGN as well, differs from a `standard' AGN because its intrinsic accretion rate is near the critical value of $\mu=0.02$, which separate the population of AGN into radiative and jetted. Therefore, any small perturbation in its accretion flow can potentially trigger the change of state in the inner part of the disk.\\For Mrk 1018, we have further explored a possible process that triggers the evolution of the accretion flow (Sec. \ref{sec:speculation}). We propose that clouds of gas might be brought close ($<$ pc) to the accretion disk by a galactic phenomenon (e.g., galactic friction \citealt{chandrasekhar43a,chandrasekhar43b,chandrasekhar43c,tremaine75,ostriker99}) and/or by an extragalactic process (a past merging event \citealt{volonteri05,hopkins06,hopkins08,kormendy13,gao20}, CCA \citealt{gaspari13,gaspari17,gaspari20,maccagni21}). When one of these gaseous structures pass near the accretion disk, as it has been observed in our own Galaxy \citep{valencia15}, it gets disrupted and its material accumulates within the disk. If the accretion rate of the cloud onto the disk is larger than the one of the disk onto the SMBH, a portion of the accretion flow puffs up, to the point that it switch into an ADAF.\\The transition establishes instabilities \citep{noda18,stern18,dexter19a,sniegowska20,feng21}, which reduce the viscous timescale to $\sim$ 10 years. Moreover, the formation of a jet \citep{rees82,narayan95,narayan96,yuan04,yuan09,taam12} produces the observed broad-band spectro-photometric variability. Especially the optical-to-X-ray dimming and the change in the X-ray hardness ratio are likely due to the disk being magnetized \citep{ferreira03,ferreira06,marcel18a,marcel18b,marcel19,marcel20}.\\We have also found in the literature some other evidences supporting the hypothesis of the state-transition of the inner accretion flow \citep{furst16,hutsem20,brogan23,walsh23}.\\Overall, our study suggests that the changing-look behaviour exhibited by Mrk 1018, and maybe by the majority of CL-AGN as well, can be attributed to galactic/extragalactic ($\gg$ kpc scale) phenomena, like mergers and/or CCA, coupled with processes happening at the level of the accretion disk (sub-pc scale), like the intrinsic critical accretion rate ($\mu\sim0.02$) and magnetic fields. These findings provide significant insights into the physics driving CL-AGN and highlight the importance of considering not only the innermost regions of an AGN, but also its wider host galaxy and even further to the extragalactic environment in order to understand the complexity of processes driving AGN activity.\\Moreover, our results are suggesting that accretion in stellar-mass black hole X-ray binaries and in AGN might be explained with the same physics, as the magnetized disk model was developed for X-ray binaries (e.g., \citealt{ferreira03}). In other words, accretion onto a black hole could be described in the same way for a stellar black hole or a SMBH, spanning about 8 orders of magnitude in mass. If correct, this will not only change the approach in the understanding of such phenomena, but it also means that we can do AGN physics on high signal-to-noise observations of nearby X-ray binaries and vice versa. 

\section{Future Prospects}\label{sec:future}
Although our conclusions are supported by observational and theoretical evidences, there are still several crucial steps that need to be taken in order to validate the proposed scenario. If the AGN is indeed undergoing inefficient accretion, a significant portion of the released energy would manifest in the form of a relativistic jet. Recent VLBI observations of Mrk 1018 \citep{walsh23} have speculated the presence of a jet but more high-resolution ($<1$ mas, i.e., sub-pc) data are needed, also to confirm that the disappearing of the broad optical lines is due to the interaction between the hypothetic jet and the surrounding BLR.\\The short-term variability observed over a span of $\sim$ 10 years suggests that the jet could have persisted for a few decades, expanding to a maximum size of 6 parsecs from the central SMBH. Assuming that this changing-look behaviour occurred throughout the entire AGN lifetime of $\sim$ $10^5$ years \citep{schawinski15}, we would expect to observe the presence of $\sim$ $10^3$-$10^4$ `mini-jets' expanding within the galaxy. The interaction of these mini-jets with the molecular gas in the host galaxy could be explored by investigating potential outflows or winds \citep{oosterloo07,morganti15,wagner16,murthy19,morganti21,maccagni21}.\\As a past wet merging event and/or CCA could be the primary factor behind the changing-look phenomenon observed in Mrk 1018, it becomes imperative to obtain high-resolution data, both spatially and spectrally, concerning the multi-phase gas distribution within and around this galaxy. While MUSE observation taken in 2015 will provide insights into the ionized gas component, our understanding of the host galaxy remains limited. To comprehensively investigate this scenario and confirm the importance of mergers and CCA for Mrk 1018, and maybe for others CL-AGN, it is essential to trace the molecular gas using the Atacama Large Millimeter/submillimeter Array (ALMA, \citealt{alma}), as well as to examine the distribution of neutral hydrogen and the radio continuum emissions with the Karoo Radio Telescope (MeerKAT, \citealt{meerkat}). These two telescopes offer the necessary angular ($\sim1''$, i.e. , $\sim1$ kpc) and spectral resolutions ($<$ 5 km/s) required for such in-depth exploration.\\The second ingredient contributing to the description of Mrk 1018 changing-look behaviour relates to its intrinsic accretion rate, which appears to be in proximity of the critical value $\mu=0.02$. If this parameter indeed plays a pivotal role in triggering the changing-look phenomenon, it stands to reason that a substantial proportion of CL-AGN would also exhibit accretion rates close to $\mu=0.02$. To ascertain the validity of this hypothesis, a rigorous statistical analysis involving a large sample of CL-AGN becomes indispensable.\\Another crucial aspect that requires attention is the development of a comprehensive model for the broad-band spectrum of an AGN with the proposed properties exhibited by Mrk 1018. This model should incorporate an outer optically thick and geometrically thin accretion disk, overlaid with a warm comptonizing medium, as well as an inner optically thin and geometrically thick flow, accompanied by a hot comptonizing medium surrounding the SMBH. Additionally, the potential presence of a jet must be considered in the model. Interestingly, a valid candidate comes again from the studies of X-ray binaries: the physically motivated Jet-Emitting Disk-Standard Accretion Disk (JED-SAD, \citealt{petrucci08,marcel18a,marcel18b}) has been tentatively applied to AGN by \citet{ursini20}, but the lack of a direct observation of the jet was limiting their interpretation. Combining this modelling with high-resolution VLBI experiments that can probe the presence of jet during this faint state of Mrk 1018, might be the missing piece of the puzzle to understand the spectral characteristics and physical processes at play in Mrk 1018 and in CL-AGN.\\The hypothesis presented in Sec. \ref{sec:speculation} is speculative and qualitative. It is imperative to develop numerical simulations of the interaction between a gaseous cloud and an accretion disk, in order to understand if our speculation is feasible and under which conditions.\\Finally, if confirmed, our proposed scenario that a large-scale ($\gg$ kpc) process, like a galactic merger and/or CCA, is a prompt for the state-transition of CL-AGN accretion disks can be used as a starting point to insert CL-AGN into the current understanding for AGN feeding and feedback \citep{ciotti07,cattaneo09,alexander12,fabian12,feng14,harrison17,morganti17}, which is still missing these rare but intriguing objects in its framework.

\begin{acknowledgements}
We thank the anonymous referee for the constructive comments and the suggested improvements for the quality and clarity of this paper.\\This research has made use of data obtained from the Chandra Data Archive and the Chandra Source Catalog, and software provided by the Chandra X-ray Center (CXC) in the application packages CIAO and Sherpa. This research was based on observations obtained with XMM-Newton, an ESA science mission with instruments and contributions directly funded by ESA Member States and NASA. We acknowledge also the use of public data from the Swift data archive.\\This research has made use of the SIMBAD database, operated at CDS, Strasbourg, France.\\SV acknowledge G. Marcel for the insightful discussion about magnetized disks and the link between AGN and X-ray binaries, and M. Trebitsch for useful comments about timescales.\\PS and CV acknowledge financial contribution from Bando Ricerca Fondamentale INAF 2022 Large Grant “Dual and binary supermassive black holes in the multi-messenger era: from galaxy mergers to gravitational waves”.\\This work has received funding from the European Research Council (ERC) under the European Union’s Horizon 2020 research and innovation programme (grant agreement No 882793 `MeerGas')
\end{acknowledgements}

\bibliographystyle{aa}
\bibliography{reference.bib}

\begin{table*}
    \caption{\label{table:obs}Main properties and statistics of the observations of Mrk 1018.}
\centering
\begin{tabular}{c c c c c c c c}
\hline\hline
Telescope & Date & ObsID & Exposure & Bkg level & Flux 0.5-2 keV & Flux 2-10 keV\\
&&& [ks] && [10\textsuperscript{-12} $\frac{erg}{s\text{ }cm^{-2}}$] & [10\textsuperscript{-12} $\frac{erg}{s\text{ }cm^{-2}}$]\\
\hline
\xmm{} & 15/01/2005 & 0201090201 & 1.61 & 4\% & 6.47$^{+0.44}_{-0.44}$ & 9.52$^{+0.66}_{-0.66}$\\
\xr{} & 29/06/2007\tablefootmark{a} & combined\tablefootmark{b} & 21.10 & 0.4\% & 7.15$^{+0.87}_{-1.96}$ & 11.73$^{+0.08}_{-0.23}$\\
\xmm{} & 07/08/2008 & 0554920301 & 12.03 & $<$ 1\% & 9.12$^{+0.38}_{-0.38}$ & 12.07$^{+0.50}_{-0.50}$\\
\ch{} & 27/11/2010 & 12868 & 25.01 & $<$ 0.1\% & 4.10$^{+0.03}_{-0.03}$ & 21.18$^{+0.14}_{-0.13}$\\
\nus{} & 10/02/2016 & 60160087002 & 21.62 & 13\% & -                    & 1.79$^{+0.11}_{-0.11}$\\
\nus{} & 05/01/2018 & 60301022002 & 27.05 & 17\% & -                    & 2.04$^{+0.03}_{-0.03}$\\
\nus{} & 05/03/2018 & 60301022003 & 43.43 & 22\% & -                    & 1.55$^{+0.16}_{-0.16}$\\
\ch{} & 05/05/2018\tablefootmark{c} & combined\tablefootmark{d} & 217.59 & $<$ 0.1\% & 0.92$^{+0.28}_{-0.01}$ & 1.99$^{+0.01}_{-0.01}$\\
\xr{} & 24/05/2018\tablefootmark{e} & combined\tablefootmark{f} & 61.10 & 1.5\% & 0.73$^{+0.07}_{-0.21}$ & 1.78$^{+0.21}_{-0.15}$\\
\nus{} & 17/07/2018 & 60301022005 & 41.84 & 17\% & -                    & 2.02$^{+0.02}_{-0.02}$\\
\xmm{} & 23/07/2018 & 0821240201 & 56.57 & 2\% & 1.10$^{+0.06}_{-0.06}$ & 2.16$^{+0.11}_{-0.11}$\\
\xmm{} & 04/01/2019 & 0821240301 & 24.19 & 3\% & 0.51$^{+0.04}_{-0.04}$ & 1.02$^{+0.09}_{-0.09}$\\
\end{tabular}
\tablefoot{The exposures of XMM are those after the filtering for the high flaring-background episodes, and are the total times of all the three cameras.\\
    \tablefoottext{a}{The date is taken as the mean of the dates of the combined data sets.}
    \tablefoottext{b}{The combined observations are: 00035166001, 00030955001, 00030955002, 00030955003, 00035776001.}
    \tablefoottext{c}{The date is taken as the mean of the dates of the combined data sets.}
    \tablefoottext{d}{The combined observations are: 18789, 19560, 20366, 20367, 20368, 20369, 20370, 21432, 22082, 21433.}
    \tablefoottext{e}{The date is taken as the mean of the dates of the combined data sets.}
    \tablefoottext{f}{The combined observations are: 00080898001, 00080898002, 00088207001, 00088207002, 00088207003, 00035776002, 00035776003, 00035776004, 00035776005, 00035776006, 00035776007, 00035776008, 00035776010, 00035776011, 00035776012, 00035776014, 00035776015, 00035776016, 00035776017, 00035776018, 00035776019, 00035776020, 00035776021, 00035776023, 00035776024, 00035776025, 00035776026, 00035776027, 00035776029, 00035776031, 00035776032, 00035776033, 00035776034, 00035776035, 00035776036, 00035776037, 00035776038, 00035776039, 00035776040, 00035776041, 00035776042, 00035776043, 00035776044, 00035776045, 00035776046, 00035776047, 00035776049, 00035776050, 00035776051, 00035776052, 00035776054, 00035776055, 00035776056, 00035776057, 00035776058, 00035776059.}
}
\end{table*}
\begin{table*}
\caption{Outcomes of the best-fit model for each available data set.}
\label{table:result}        
\centering          
\begin{tabular}{c c c c c c c c c}
\hline\hline     
Telescope & ObsID & kT & $\Gamma_{WC}$ & $k_{WC}$ & $\Gamma$ & $k$ & $I$ & $\chi^2_{red}$ \\
&& [keV] && [10\textsuperscript{-3} $\frac{photons}{keV\text{ }s\text{ }cm^{-2}}$] && [10\textsuperscript{-3} $\frac{photons}{keV\text{ }s\text{ }cm^{-2}}$] & [10\textsuperscript{-6} $\frac{photons}{keV\text{ }s\text{ }cm^{-2}}$] &\\
\hline                    
\xmm{} & 0201090201 &
0.46$^{+0.17}_{-0.09}$ & 2.69$^{+0.13}_{-0.13}$ & 1.11$^{+0.05}_{-0.05}$ &
1.62$^{+0.13}_{-0.13}$ & 2.26$^{+0.41}_{-0.41}$ &
9.5$^{+7.7}_{-7.7}$ & 0.96 \\
\xr{} & combined &
0.70$^{+0.00}_{-0.37}$ & 2.82$^{+0.18}_{-0.28}$ & 0.92$^{+0.10}_{-0.09}$ &
1.62$^{+0.23}_{-0.22}$ & 2.65$^{+1.10}_{-0.77}$ &
$<19.0$ & 0.95 \\
\xmm{} & 0554920301 &
0.43$^{+0.04}_{-0.03}$ & 2.60$^{+0.05}_{-0.05}$ & 1.60$^{+0.03}_{-0.03}$ &
1.69$^{+0.05}_{-0.05}$ & 3.14$^{+0.24}_{-0.22}$ &
10.4$^{+3.3}_{-3.3}$ & 1.10 \\
\ch{} & 12868 &
0.26$^{+0.03}_{-0.03}$ & 1.62$^{+0.20}_{-0.23}$ & 1.66$^{+0.26}_{-0.26}$ &
1.88$^{+0.11}_{-0.13}$ & 4.78$^{+0.82}_{-0.75}$ &
$<17.0$ & 1.22 \\
\nus{} & 60160087001 &
- & - & - &
1.77$^{+0.11}_{-0.11}$ & 0.49$^{+0.13}_{-0.11}$ &
6.1$^{+3.1}_{-3.1}$ & 0.67\\
\nus{} & 60160087002 &
- & - & - &
1.69$^{+0.08}_{-0.08}$ & 0.51$^{+0.10}_{-0.09}$ &
3.11$^{+2.5}_{-2.5}$ & 1.11 \\
\nus{} & 60301022003 &
- & - & - &
1.73$^{+0.08}_{-0.08}$ & 0.40$^{+0.07}_{-0.06}$ &
1.06$^{+1.7}_{-1.1}$ & 1.00 \\
\ch{} & combined &
0.19$^{+0.38}_{-0.05}$ & $<3.42$ & 0.03$^{+0.01}_{-0.01}$ &
1.60$^{+0.01}_{-0.01}$ & 0.44$^{+0.01}_{-0.01}$ &
3.5$^{+0.9}_{-0.9}$ & 1.07\\
\xr{} & combined &
0.21 (fix) & 1.85 (fix) & $<0.03$ &
1.58$^{+0.05}_{-0.10}$ & 0.37$^{+0.01}_{-0.01}$ &
7.0$^{+6.1}_{-6.4}$ & 0.89\\
\nus{} & 60301022005 &
- & - & - &
1.69$^{+0.07}_{-0.07}$ & 0.51$^{+0.08}_{-0.07}$ &
3.6$^{+2.2}_{-2.2}$ & 1.09\\
\xmm{} & 0821240201 &
0.21$^{+0.02}_{-0.02}$ & 1.85$^{+0.21}_{-0.22}$ & 0.07$^{+0.01}_{-0.01}$ &
1.62$^{+0.05}_{-0.05}$ & 0.52$^{+0.05}_{-0.04}$ &
6.6$^{+0.8}_{-0.8}$ & 1.09\\
\xmm{} & 0821240301 &
0.14$^{+0.03}_{-0.02}$ & $<1.68$ & 0.07$^{+0.01}_{-0.01}$ &
1.63$^{+0.12}_{-0.12}$ & 0.26$^{+0.05}_{-0.04}$ &
2.9$^{+1.0}_{-1.0}$ & 1.02\\
\hline                  
\end{tabular}
\tablefoot{$kT$ is the warm corona temperature, $\Gamma_{WC}$ the Warm Comptonization power law spectral index, $k_{WC}$ the Warm Comptonization power law normalization, $\Gamma$ the primary power law photon index related to the hot Comptonization, $k$ the primary power law normalization and $I$ the Fe K$\alpha$ emission line intensity.}
\end{table*}
\longtab{
\begin{longtable}{c c c c c c c c}
    \caption{\label{table:om2} Optical/UV flux densities of Mrk 1018 in unit of 10\textsuperscript{-16} erg s\textsuperscript{-1} cm\textsuperscript{-2} \AA\textsuperscript{-1} for each filter. N.A. = Not Available.}\\ 
    \hline\hline 
Telescope & ObsID & V & B & U & UVW1 & UVM2 & UVW2 \\  
    \hline
    \endfirsthead
    \caption{continued.}\\
    \hline\hline
Telescope & ObsID & V & B & U & UVW1 & UVM2 & UVW2 \\
    \hline   
    \endhead
    \hline
    \endfoot
\om{} & 0201090201 & N.A & N.A & 69.6 $\pm$ 0.5 & 95.0 $\pm$ 0.6 & 118 $\pm$ 1 & 137 $\pm$ 2 \\
\uv{} & 00035166001 & 59.1 $\pm$ 4.6 & 57.7 $\pm$ 4.1 & 87.0 $\pm$ 6.0 & 135 $\pm$ 7 & 153 $\pm$ 8 & 207 $\pm$ 8 \\
\uv{} & 00030955001 & N.A. & N.A. & N.A. & 119 $\pm$ 3 & N.A. & N.A. \\
\uv{} & 00030955002 & N.A. & N.A. & N.A. & N.A. & N.A. & 164 $\pm$ 3 \\
\uv{} & 00030955003 & N.A. & N.A. & N.A. & 122 $\pm$ 3 & N.A. & N.A. \\
\uv{} & 00035776001 & 54.7 $\pm$ 2.0 & 54.8 $\pm$ 1.6 & 70.2 $\pm$ 2.1 & 102 $\pm$ 3 & 110 $\pm$ 3 & 137 $\pm$ 3 \\
\om{} & 0554920301 & N.A & N.A & 75.2 $\pm$ 0.2 & 99.2 $\pm$ 0.4 & 227 $\pm$ 1 & N.A. \\
\uv{} & 00049654001 & 38.1 $\pm$ 1.9 & 25.7 $\pm$ 1.2 & 16.3 $\pm$ 0.9 & 16.3 $\pm$ 1.0 & 17.1 $\pm$ 1.0 & 16.4 $\pm$ 0.8 \\
\uv{} & 00049654002 & 37.2 $\pm$ 1.8 & 28.3 $\pm$ 1.2 & 19.7 $\pm$ 1.0 & 23.8 $\pm$ 1.2 & 20.9 $\pm$ 1.7 & 25.6 $\pm$ 1.0 \\
\uv{} & 00080898001 & 38.9 $\pm$ 1.5 & 24.9 $\pm$ 0.9 & 11.5 $\pm$ 0.6 & 7.5 $\pm$ 0.6 & 5.1 $\pm$ 0.5 & 5.6 $\pm$ 0.4 \\
\uv{} & 00080898002 & 38.2 $\pm$ 1.5 & 25.0 $\pm$ 1.0 & 9.7 $\pm$ 0.6 & 7.4 $\pm$ 0.6 & 5.1 $\pm$ 0.3 & 6.4 $\pm$ 0.4 \\
\uv{} & 00088207001 & N.A. & N.A. & 10.6 $\pm$ 0.3 & N.A. & N.A. & N.A. \\
\uv{} & 00088207002 & N.A. & N.A. & N.A. & N.A. & N.A. & 7.06 $\pm$ 0.2 \\
\uv{} & 00088207003 & N.A. & N.A. & 10.8 $\pm$ 0.3 & N.A. & N.A. & N.A. \\
\om{} & 0821240201 & N.A & N.A & N.A. & N.A. & 0.71 $\pm$ 0.09 & N.A. \\
\uv{} & 00035776002 & 37.2 $\pm$ 1.6 & 23.9 $\pm$ 0.9 & 11.3 $\pm$ 0.6 & 7.8 $\pm$ 0.6 & 5.0 $\pm$ 0.5 & 6.5 $\pm$ 0.5 \\
\uv{} & 00035776003 & 37.6 $\pm$ 1.6 & 24.3 $\pm$ 0.9 & 10.1 $\pm$ 0.6 & 6.8 $\pm$ 0.6 & 4.6 $\pm$ 0.5 & 6.8 $\pm$ 0.5 \\
\uv{} & 00035776004 & 37.7 $\pm$ 1.6 & 23.8 $\pm$ 0.9 & 11.3 $\pm$ 0.6 & 7.1 $\pm$ 0.6 & 4.7 $\pm$ 0.5 & 6.2 $\pm$ 0.4 \\
\uv{} & 00035776005 & 34.3 $\pm$ 2.0 & 23.3 $\pm$ 1.2 &  8.1 $\pm$ 0.7 & 7.9 $\pm$ 0.8 & 5.4 $\pm$ 0.7 & 6.4 $\pm$ 0.6 \\
\uv{} & 00035776006 & 37.7 $\pm$ 1.7 & 23.8 $\pm$ 1.0 & 10.5 $\pm$ 0.7 & 6.5 $\pm$ 0.6 & 4.6 $\pm$ 0.6 & 6.4 $\pm$ 0.5 \\
\uv{} & 00035776007 & 36.0 $\pm$ 1.5 & 25.4 $\pm$ 1.0 & 10.5 $\pm$ 0.6 & 6.9 $\pm$ 0.6 & 4.2 $\pm$ 0.5 & 5.0 $\pm$ 0.4 \\
\uv{} & 00035776008 & 37.3 $\pm$ 2.0 & 25.3 $\pm$ 1.2 &  9.9 $\pm$ 0.8 & 8.2 $\pm$ 0.8 & 4.4 $\pm$ 0.7 & 6.8 $\pm$ 0.6 \\
\uv{} & 00035776010 & 36.9 $\pm$ 1.6 & 23.6 $\pm$ 0.9 & 10.9 $\pm$ 0.6 & 7.4 $\pm$ 0.6 & 4.4 $\pm$ 0.5 & 6.3 $\pm$ 0.5 \\
\uv{} & 00035776011 & 35.9 $\pm$ 2.5 & 24.5 $\pm$ 0.9 & 11.1 $\pm$ 0.6 & 6.8 $\pm$ 0.6 & 4.6 $\pm$ 0.5 & 7.1 $\pm$ 0.5 \\
\uv{} & 00035776012 & 36.1 $\pm$ 2.0 & 24.6 $\pm$ 1.2 & 10.9 $\pm$ 0.8 & 6.3 $\pm$ 0.7 & 4.2 $\pm$ 0.7 & 6.1 $\pm$ 0.6 \\
\uv{} & 00035776014 & 37.2 $\pm$ 1.6 & 24.3 $\pm$ 0.9 & 10.0 $\pm$ 0.6 & 7.0 $\pm$ 0.6 & 4.8 $\pm$ 0.5 & 6.3 $\pm$ 0.5 \\
\uv{} & 00035776015 & 37.0 $\pm$ 1.6 & 24.6 $\pm$ 0.9 & 10.8 $\pm$ 0.6 & 6.7 $\pm$ 0.6 & 4.9 $\pm$ 0.5 & 6.1 $\pm$ 0.4 \\
\uv{} & 00035776016 & 35.1 $\pm$ 1.6 & 24.0 $\pm$ 1.0 & 10.0 $\pm$ 0.6 & 7.1 $\pm$ 0.6 & 5.5 $\pm$ 0.6 & 6.3 $\pm$ 0.5 \\
\uv{} & 00035776017 & 37.7 $\pm$ 1.9 & 25.6 $\pm$ 1.2 & 10.6 $\pm$ 0.8 & 8.0 $\pm$ 0.8 & 4.4 $\pm$ 0.7 & 6.6 $\pm$ 0.6 \\
\uv{} & 00035776018 & 36.9 $\pm$ 1.5 & 24.5 $\pm$ 0.9 & 10.3 $\pm$ 0.6 & 8.0 $\pm$ 0.6 & 4.8 $\pm$ 0.5 & 6.0 $\pm$ 0.4 \\
\uv{} & 00035776019 & 38.1 $\pm$ 2.1 & 24.1 $\pm$ 1.3 & 11.6 $\pm$ 0.9 & 8.8 $\pm$ 0.9 & 4.6 $\pm$ 0.7 & 6.7 $\pm$ 0.7 \\
\uv{} & 00035776020 & 36.4 $\pm$ 1.5 & 23.5 $\pm$ 0.9 & 10.8 $\pm$ 0.6 & 8.5 $\pm$ 0.6 & 4.5 $\pm$ 0.4 & 6.4 $\pm$ 0.4 \\
\uv{} & 00035776021 & 37.0 $\pm$ 2.1 & 24.4 $\pm$ 1.3 & 10.3 $\pm$ 0.8 & 8.9 $\pm$ 0.9 & 5.8 $\pm$ 0.8 & 6.3 $\pm$ 0.6 \\
\uv{} & 00035776023 & 40.8 $\pm$ 1.7 & 25.0 $\pm$ 1.0 & 11.5 $\pm$ 0.7 & 7.0 $\pm$ 0.6 & 4.3 $\pm$ 0.5 & 6.5 $\pm$ 0.5 \\
\uv{} & 00035776024 & 35.8 $\pm$ 2.1 & 24.8 $\pm$ 1.3 & 11.3 $\pm$ 0.9 & 8.4 $\pm$ 0.9 & 4.5 $\pm$ 0.7 & 6.7 $\pm$ 0.7 \\
\uv{} & 00035776025 & 36.2 $\pm$ 1.5 & 24.4 $\pm$ 0.9 & 10.9 $\pm$ 0.6 & 7.4 $\pm$ 0.6 & 4.3 $\pm$ 0.5 & 5.4 $\pm$ 0.4 \\
\uv{} & 00035776026 & 37.3 $\pm$ 1.7 & 25.7 $\pm$ 1.1 & 10.9 $\pm$ 0.7 & 8.0 $\pm$ 0.7 & 4.9 $\pm$ 0.6 & 6.9 $\pm$ 0.5 \\
\uv{} & 00035776027 & 35.9 $\pm$ 1.5 & 23.9 $\pm$ 0.9 & 10.4 $\pm$ 0.6 & 7.2 $\pm$ 0.6 & 6.0 $\pm$ 0.6 & 6.4 $\pm$ 0.4 \\
\uv{} & 00035776029 & 37.1 $\pm$ 1.5 & 23.6 $\pm$ 0.9 &  9.7 $\pm$ 0.6 & 7.5 $\pm$ 0.5 & 5.3 $\pm$ 0.5 & 6.2 $\pm$ 0.4 \\
\uv{} & 00035776031 & 36.5 $\pm$ 1.6 & 25.2 $\pm$ 1.0 &  9.7 $\pm$ 0.6 & 7.9 $\pm$ 0.6 & 4.6 $\pm$ 0.5 & 6.8 $\pm$ 0.5 \\
\uv{} & 00035776032 & 41.2 $\pm$ 2.3 & 25.7 $\pm$ 1.3 &  9.4 $\pm$ 0.8 & 6.4 $\pm$ 0.8 & 4.7 $\pm$ 0.8 & 5.3 $\pm$ 0.6 \\
\uv{} & 00035776033 & 36.8 $\pm$ 1.5 & 24.8 $\pm$ 0.9 & 10.4 $\pm$ 0.6 & 7.0 $\pm$ 0.6 & 4.7 $\pm$ 0.5 & 6.5 $\pm$ 0.4 \\
\uv{} & 00035776034 & 35.6 $\pm$ 1.5 & 23.8 $\pm$ 0.9 & 10.5 $\pm$ 0.6 & 6.9 $\pm$ 0.5 & 4.1 $\pm$ 0.4 & 5.5 $\pm$ 0.4 \\
\uv{} & 00035776035 & 36.6 $\pm$ 1.5 & 25.7 $\pm$ 0.9 & 10.3 $\pm$ 0.6 & 7.2 $\pm$ 0.5 & 3.9 $\pm$ 0.4 & 6.0 $\pm$ 0.4 \\
\uv{} & 00035776036 & 36.8 $\pm$ 1.6 & 23.1 $\pm$ 0.9 & 10.9 $\pm$ 0.6 & 6.2 $\pm$ 0.5 & 4.2 $\pm$ 0.5 & 5.8 $\pm$ 0.4 \\
\uv{} & 00035776037 & 36.9 $\pm$ 1.5 & 24.2 $\pm$ 0.9 & 10.1 $\pm$ 0.6 & 6.9 $\pm$ 0.5 & 4.6 $\pm$ 0.5 & 5.9 $\pm$ 0.4 \\
\uv{} & 00035776038 & 38.3 $\pm$ 1.6 & 24.8 $\pm$ 0.9 &  9.3 $\pm$ 0.6 & 6.6 $\pm$ 0.5 & 5.0 $\pm$ 0.6 & 5.7 $\pm$ 0.4 \\
\uv{} & 00035776039 & 38.8 $\pm$ 1.6 & 25.6 $\pm$ 0.9 & 10.5 $\pm$ 0.6 & 6.1 $\pm$ 0.5 & 4.2 $\pm$ 0.5 & 5.7 $\pm$ 0.4 \\
\uv{} & 00035776040 & 36.2 $\pm$ 1.5 & 24.4 $\pm$ 0.9 &  9.5 $\pm$ 0.5 & 6.4 $\pm$ 0.5 & 5.0 $\pm$ 0.5 & 5.6 $\pm$ 0.4 \\
\uv{} & 00035776041 & 36.9 $\pm$ 1.6 & 24.8 $\pm$ 0.9 & 11.2 $\pm$ 0.6 & 6.1 $\pm$ 0.5 & 4.3 $\pm$ 0.5 & 5.9 $\pm$ 0.4 \\
\uv{} & 00035776042 & 37.3 $\pm$ 1.9 & 25.6 $\pm$ 1.2 & 10.1 $\pm$ 0.7 & 6.7 $\pm$ 0.7 & 5.0 $\pm$ 0.6 & 5.4 $\pm$ 0.5 \\
\uv{} & 00035776043 & 36.9 $\pm$ 1.6 & 24.7 $\pm$ 0.9 &  9.7 $\pm$ 0.6 & 6.7 $\pm$ 0.6 & 5.0 $\pm$ 0.6 & 6.3 $\pm$ 0.5 \\
\uv{} & 00035776044 & 38.7 $\pm$ 1.6 & 26.3 $\pm$ 1.0 & 10.2 $\pm$ 0.6 & 6.3 $\pm$ 0.5 & 4.2 $\pm$ 0.5 & 5.7 $\pm$ 0.4 \\
\uv{} & 00035776045 & 35.4 $\pm$ 1.4 & 25.6 $\pm$ 0.9 & 10.3 $\pm$ 0.6 & 7.7 $\pm$ 0.6 & 4.2 $\pm$ 0.5 & 6.0 $\pm$ 0.4 \\
\uv{} & 00035776046 & 34.6 $\pm$ 1.9 & 23.8 $\pm$ 1.2 & 11.2 $\pm$ 0.8 & 6.6 $\pm$ 0.7 & 5.4 $\pm$ 0.8 & 5.5 $\pm$ 0.6 \\
\uv{} & 00035776047 & 36.0 $\pm$ 1.5 & 24.6 $\pm$ 0.9 & 10.2 $\pm$ 0.6 & 7.1 $\pm$ 0.6 & 4.3 $\pm$ 0.5 & 6.1 $\pm$ 0.4 \\
\uv{} & 00035776049 & 36.7 $\pm$ 1.8 & 26.3 $\pm$ 1.2 & 10.3 $\pm$ 0.7 & 7.1 $\pm$ 0.7 & 5.2 $\pm$ 0.6 & 5.9 $\pm$ 0.5 \\
\uv{} & 00035776050 & 36.8 $\pm$ 1.5 & 24.2 $\pm$ 0.9 &  9.6 $\pm$ 0.6 & 6.8 $\pm$ 0.5 & 5.4 $\pm$ 0.5 & 6.5 $\pm$ 0.5 \\
\uv{} & 00035776051 & 35.9 $\pm$ 1.7 & 23.7 $\pm$ 1.0 & 10.1 $\pm$ 0.7 & 5.9 $\pm$ 0.6 & 4.2 $\pm$ 0.6 & 5.8 $\pm$ 0.5 \\
\uv{} & 00035776052 & 35.1 $\pm$ 2.0 & 26.0 $\pm$ 1.3 & 10.7 $\pm$ 0.8 & 7.9 $\pm$ 0.8 & 5.1 $\pm$ 0.7 & 5.5 $\pm$ 0.6 \\
\uv{} & 00035776054 & 36.7 $\pm$ 1.5 & 24.8 $\pm$ 0.9 & 10.9 $\pm$ 0.6 & 7.5 $\pm$ 0.6 & 4.4 $\pm$ 0.5 & 6.2 $\pm$ 0.4 \\
\uv{} & 00035776055 & 37.2 $\pm$ 1.9 & 23.9 $\pm$ 1.2 & 10.0 $\pm$ 0.7 & 6.7 $\pm$ 0.7 & 4.3 $\pm$ 0.5 & 6.8 $\pm$ 0.6 \\
\uv{} & 00035776056 & 37.9 $\pm$ 1.5 & 24.4 $\pm$ 0.9 &  9.9 $\pm$ 0.6 & 7.2 $\pm$ 0.6 & 3.9 $\pm$ 0.5 & 5.8 $\pm$ 0.4 \\
\uv{} & 00035776057 & 38.4 $\pm$ 1.6 & 25.8 $\pm$ 1.0 & 10.5 $\pm$ 0.6 & 6.4 $\pm$ 0.5 & 4.6 $\pm$ 0.5 & 6.4 $\pm$ 0.4 \\
\uv{} & 00035776058 & 36.4 $\pm$ 1.5 & 24.9 $\pm$ 0.9 & 10.9 $\pm$ 0.6 & 6.2 $\pm$ 0.5 & 4.1 $\pm$ 0.5 & 4.8 $\pm$ 0.4 \\
\om{} & 0821240301 & N.A & N.A & N.A. & N.A. & 0.49 $\pm$ 0.16 & N.A. \\
\uv{} & 00035776059 & 38.1 $\pm$ 1.4 & 24.3 $\pm$ 0.8 & 9.8 $\pm$ 0.5 & 6.0 $\pm$ 0.4 & 3.1 $\pm$ 0.3 & 4.5 $\pm$ 0.3
\end{longtable}}
\begin{table*}
    \caption{Mrk 1018 observed and expected value for the optical/UV to X-ray flux ratio parameter $\alpha_{ox}$.}   
    \label{table:aox}      
    \centering          
    \begin{tabular}{c c c c}
    \hline\hline       
Telescope & ObsID & $\alpha_{ox}$ observed & $\alpha_{ox}$ expected \\ 
    \hline                    
\om{} & 0201090201 & -1.39$^{+0.03}_{-0.03}$ & -1.27$^{+0.13}_{-0.13}$ \\
\uv{} & combined & -1.37$^{+0.07}_{-0.05}$ & -1.27$^{+0.13}_{-0.13}$ \\
\om{} & 0554920301 & -1.45$^{+0.01}_{-0.01}$ & -1.31$^{+0.13}_{-0.13}$ \\
\uv{} & 00049654001 & -1.26$^{+0.06}_{-0.06}$ & -1.13$^{+0.13}_{-0.13}$ \\
\uv{} & 00049654002 & -1.15$^{+0.03}_{-0.03}$ & -1.15$^{+0.13}_{-0.13}$ \\
\om{} & 0821240201 & -1.12$^{+0.02}_{-0.02}$ & -1.06$^{+0.13}_{-0.13}$ \\
\uv{} & combined & -1.14$^{+0.01}_{-0.02}$ & -1.05$^{+0.13}_{-0.13}$ \\
\om{} & 0821240301 & -1.19$^{+0.03}_{-0.03}$ & -1.04$^{+0.13}_{-0.13}$ \\
    \hline                  
    \end{tabular}
    \tablefoot{The uncertainties for the expected value \citep{lusso10} were calculated at the 1$\sigma$ level using the propagation of errors formula on equation \ref{eq:theo}. The \uv{} dataset without an ObsID are those from the stacked observations.}
\end{table*}
\begin{table*}
    \caption{Best-fit result for the model \textit{AGNSED}.}   
    \label{table:agnsed}      
    \centering          
    \begin{tabular}{c c c c c c c c c c}
    \hline\hline       
State & SMBH mass & $\mu$ & $kT_{WC}$ & $\Gamma_{WC}$ & $R_{WC}$ & $\Gamma_{HC}$ & $R_{HC}$ & $\chi^2_{red}$\\
 & [$10^7 M_\odot$] & & [kev] & & [$R_g$] & & [$R_g$] &\\
    \hline                    
bright & 9.7$^{+0.7}_{-0.7}$ & 0.057$^{+0.008}_{-0.004}$ & 0.46$^{+0.15}_{-0.16}$ & 2.64$^{+0.06}_{-0.06}$ & 52$^{+6}_{-6}$ & 1.71$^{+0.06}_{-0.07}$ & 18.3$^{+1.2}_{-0.8}$ & 1.08\\
faint & 9.7 (fix) & 0.011$^{+0.001}_{-0.001}$ & 0.16$^{+0.01}_{-0.01}$ & 1.61$^{+0.01}_{-0.01}$ & 19.9$^{+0.1}_{-0.1}$ & 1.71 (fix) & 18.3 (fix) & 1.06\\
    \hline                  
    \end{tabular}
    \tablefoot{$\mu$ is the accretion rate normalized to the Eddington ratio $\frac{\dot{M}}{\dot{M}_{edd}}$, $kT_{WC}$ is the warm corona temperature in keV, $\Gamma_{WC}$ the warm Comptonization spectral index, $R_{WC}$ the warm corona size in gravitational radii ($R_g=\frac{GM}{c^2}$), $\Gamma_{HC}$ the hot Comptonization spectral index, $R_{HC}$ the hot corona size in $R_g$ and $\chi^2_{red}$ the ratio between the $\chi^2$ and the degrees of freedom.}
\end{table*}
\end{document}